\documentclass[11pt,reqno]{amsart}
\usepackage{amsmath, amssymb, amsthm}
\usepackage{graphicx, xspace}

\numberwithin{equation}{section}

\newtheorem{theorem}{Theorem}[section]
\newtheorem{definition}[theorem]{Definition}
\newtheorem{proposition}[theorem]{Proposition}
\newtheorem{corollary}[theorem]{Corollary}
\newtheorem{lemma}[theorem]{Lemma}

\def \proof {\noindent {\bf Proof.}\ \ }

\def \endproof {{\mbox{}\nolinebreak\hfill\rule{2mm}{2mm}\par\medbreak}}

\newlength{\algorithmwidth}
\DeclareMathOperator{\dist}{dist}

\DeclareMathOperator{\conv}{\triangle}
\DeclareMathOperator*{\Conv}{conv}
\DeclareMathOperator*{\aff}{aff}

\DeclareMathOperator{\facet}{facet}
\DeclareMathOperator{\Facet}{Facet}
\DeclareMathOperator*{\Span}{span}
\DeclareMathOperator*{\cone}{cone}
\DeclareMathOperator{\edges}{edges}
\DeclareMathOperator{\Recess}{Recess}
\DeclareMathOperator{\ang}{ang}
\DeclareMathOperator{\Exp}{Exp}
\DeclareMathOperator*{\argmax}{argmax}
\DeclareMathOperator{\Dom}{Dom}
\DeclareMathOperator{\Vol}{Vol}

\def \R {\mathbb{R}}

\def \E {\mathbb{E}}

\def \P {\mathbb{P}}
\def \T {\mathbb{T}}

\def \EE {\mathcal{E}}

\def \II {\mathcal{I}}
\def \MM {\mathcal{M}}

\def \PP {\mathcal{P}}

\def \e {\varepsilon}

\def \l {\lambda}
\def \s {\sigma}
\def \w {\omega}
\def \< {\langle}
\def \> {\rangle}

\def \one {\mathbf{1}}
\def \LP1 {LP$_1$}

\def \phaseI {phase\nobreakdash-I\xspace}
\def \phaseII {phase\nobreakdash-II\xspace}
\def \UnitLP {\mbox{Unit LP\xspace}}
\def \IntLP {\mbox{Int LP\xspace}}
\def \LP {\mbox{LP\xspace}}

\begin{document}

\title[]{Beyond Hirsch Conjecture: walks on random polytopes
  and smoothed complexity of the simplex method}

\author{Roman Vershynin}
\address{Department of Mathematics, University of California, Davis, CA 95616, U.S.A.}
\email{vershynin@math.ucdavis.edu}

\thanks{Partially supported by NSF grants DMS 0401032 and 0652617
  and Alfred P. Sloan Foundation}

\begin{abstract}
  The smoothed analysis of algorithms is concerned with the expected running
  time of an algorithm under slight random perturbations of arbitrary inputs.
  Spielman and Teng proved that the shadow-vertex simplex method has polynomial
  smoothed complexity. On a slight random perturbation
  of an arbitrary linear program, the simplex method finds the solution
  after a walk on polytope(s) with expected length polynomial in the number
  of constraints $n$, the number of variables $d$ and the inverse
  standard deviation of the perturbation $1/\sigma$.

  We show that the length of walk in the simplex method is
  actually {\em polylogarithmic} in the number of constraints $n$.
  Spielman-Teng's bound on the walk was
  $O^*(n^{86} d^{55} \s^{-30})$, up to logarithmic factors.
  We improve this to $O(\log^7 n (d^9 + d^3 \s^{-4}))$.
  This shows that the tight Hirsch conjecture $n-d$ on the length of walk
  on polytopes is not a limitation for the smoothed Linear Programming.
  Random perturbations create short paths between vertices.

  We propose a randomized \phaseI for solving arbitrary linear programs,
  which is of independent interest.
  Instead of finding a vertex of a feasible set, we add a vertex at random to
  the feasible set. This does not affect the solution of the linear program with
  constant probability. So, in expectation it takes a constant number of independent
  trials until a correct solution is found. This overcomes
  one of the major difficulties of smoothed analysis of the simplex method
  -- one can now statistically decouple the walk from the smoothed linear program.
  This yields a much better reduction of the smoothed complexity to a geometric
  quantity -- the size of planar sections of random polytopes.
  We also improve upon the known estimates for that size, showing that it is
  polylogarithmic in the number of vertices.
\end{abstract}

\maketitle

\setcounter{tocdepth}{1}
\tableofcontents

\section{Introduction}

The simplex method is ``the classic example of an algorithm that is known to
perform well in practice but which takes exponential time in the worst case''
\cite{ST}. In an attempt to explain this behavior, Spielman and Teng \cite{ST}
introduced the concept of {\em smoothed analysis} of algorithms,
in which one measures the expected complexity of an algorithm
under slight random perturbations of arbitrary inputs. They proved that a
variant of the simplex method has polynomial smoothed complexity.

Consider a linear program of the form
\begin{equation*}                       \tag{LP}
\begin{aligned}
  &\text{maximize } \< z, x \>  \\
  &\text{subject to } Ax \le b,
\end{aligned}
\end{equation*}
where $A$ is an $n \times d$ matrix, representing $n$ constraints,
and $x$ is a vector representing $d$ variables.

A simplex method starts at some vertex $x_0$ of the polytope $Ax \le b$, found
by a \phaseI method, and then walks on the vertices of the polytope toward the
solution of (LP). A pivot rule dictates how to choose the next vertex in this walk.
The complexity of the simplex method is then determined by the length of the walk --
the number of pivot steps.

So far, smoothed analysis has only been done for
the shadow-vertex pivot rule introduced by Gaas and Saaty \cite{GS}.
The shadow-vertex simplex method first chooses an {\em initial objective function}
$z_0$ optimized by the initial vertex $x_0$.
Then it interpolates between $z_0$ and the actual objective function $z$.
Namely, it rotates $z_0$ toward $z$ and computes the vertices that optimize all
the objective functions between $z_0$ and $z$.

A {\em smoothed linear program} is a linear program of the form (LP),
where the rows $a_i$ of $A$, called the {\em constraint vectors},
and $b$ are independent Gaussian random vectors, with arbitrary
centers $\bar{a}_i$ and $\bar{b}$ respectively, and whose coordinates
are independent normal random variables with standard deviations
$\s \max_i \|(\bar{a}_i, \bar{b}_i)\|$.
Spielman and Teng proved

\begin{theorem} \cite{ST}                   \label{Spielman-Teng}
  For arbitrary linear program with $d>3$ variables and $n>d$ constraints,
  the expected number of pivot steps
  in a two-phase shadow-vertex simplex method for the smoothed program
  is at most a polynomial $\PP(n,d,\s^{-1})$.
\end{theorem}

\noindent Spielman-Teng's analysis yields the following estimate
on expected number of pivot steps:
$$
\PP(n,d,\s^{-1}) \le O^*(n^{86} d^{55} \s^{-30})
$$
where the logarithmic factors are disregarded. The subsequent work of Deshpande
and Spielman \cite{DS} improved on the exponents of $d$ and $\s$; however, it
doubled the exponent of $n$.

Another model of randomness, in which the directions of the inequalities
in (LP) are chosen independently at random, was studied in the eighties
\cite{Hai83, Adl83, Tod86, AM85, AKS87}. It was shown that a shadow-vertex
simplex method solves such problems in $O(d^2)$ expected number of pivot steps.
Note that this bound does not depend on the number of constraints $n$.
However, it is not clear whether reversion of the inequalities can be
justified as a good model for typical linear problems.

These results lead to the question -- how does the smoothed complexity
of the simplex method depend on the number of constraints $n$?
Unlike in the model with randomly reversed inequalities,
the number of pivots steps in the smoothed analysis must be at least
logarithmic in $n$ (see below).

In this paper, we prove the first polylogarithmic upper bound:

\begin{theorem}[Main]                       \label{main}
  The expected number of pivot steps in Theorem~\ref{Spielman-Teng} is at most
  $$
  \PP(n,d,\s^{-1}) \le O(\log^7 n (d^9 + d^3 \s^{-4})).
  $$
 \end{theorem}
\noindent See Theorem~\ref{main precise} for a more precise estimate.

\medskip

So, the number of pivot steps is {\em polylogarithmic in the number
of constraints} $n$, while the previous known bounds were polynomial in $n$.

This bound goes in some sense beyond the classical Hirsch conjecture on the
diameter of polytopes, the maximal number of steps in the shortest walk
between any pair of vertices (see e.g. \cite{KK92}).
Hirsch conjecture states that the diameter of a polytope with $n$ faces
in $d$ dimensions, and in particular of the feasible polytope of (LP),
is at most $n-d$.

Hirsch conjecture is tight, so it would be natural to think of $n-d$ as a lower
bound on the worst case complexity of {\em any} variant of the simplex method.
Theorem~\ref{main} (and Theorem~\ref{section} below) claim that
a random perturbation creates a much shorter path between
a given pair of vertices.
Moreover, while Hirsch conjecture does not suggest any algorithm for finding
a short walk, the shadow-vertex simplex method already finds a much shorter walk!

\medskip

One can wonder whether a random perturbation creates a short path between the vertices by destroying most of them. This is not so even in the average case,
when $A$ is a matrix with i.i.d. Gaussian entries.
Indeed, the expected number of vertices of the random polytope
is asymptotic to
$2^d d^{-1/2} (d-1)^{-1} (\pi \log n)^{(d-1)/2}$ (\cite{R}, see \cite{H}).
This bound is exponential in $d$ and sublinear but not polylogarithmic in $n$,
as Theorem~\ref{main} claims.

For $d=2$, the random polygon will have $\Omega(\sqrt{\log n})$ vertices,
which is thus a lower bound on the number of pivot steps of (any) simplex
algorithm.

\medskip

The smoothed complexity (expected running time) of the simplex method
is $O(\PP(n,d,\s^{-1}) \; t_\text{pivot})$,
where $t_\text{pivot}$ is the time to make one pivot step under the shadow-vertex
pivot rule. The dependence of $t_\text{pivot}$ on $n$ is at most linear,
for one only needs to find an appropriate vector $a_i$ among the $n$ vectors
to update the running vertex. However, for many well structured linear problems
the exhaustive search over all $a_i$ is not necessary, which makes
$t_\text{pivot}$ much smaller. In these cases, Theorem~\ref{main} shows
that {\em the shadow-vertex simplex method can solve very large scale problems}
(up to exponentially many constraints).

\subsection*{Acknowledgements}
The author is grateful to the referees for careful reading of the manuscript
and many useful remarks and suggestions, which greatly improved the paper.

\section{Outline of the approach}

Our smoothed analysis of the simplex method is largely inspired
by that of Spielman and Teng \cite{ST}. We resolve a few conceptual
difficulties that arise in \cite{ST}, which eventually simplifies
and improves the overall picture.

\subsection{Interpolation: reduction to unit linear programs}

First, we reduce an arbitrary linear program (LP) to a {\em unit linear program} --
the one for which $b = \one$. This is done by a simple interpolation.
This observation is independent of any particular algorithm to solve linear programs.

An interpolation variable will be introduced,
and (LP) will reduce to a unit program in dimension $d+1$
with constraint vectors of type $(a_i,b_i)$.  A simple but very useful consequence
is that this reduction preserves the Gaussian distribution of the constraints --
if (LP) has independent Gaussian constraints (as the smoothed program does),
then so does the reduced unit program.

\subsection{Duality: reduction to planar sections of random polytopes}  \label{s:duality}

Now that we have a unit linear program, it is best viewed in the polar perspective.
The polar of the feasible set $Ax \le \one$ is the polytope
$$
P = \Conv(0, a_1,\ldots,a_n).
$$
The unit linear problem is then equivalent to finding $\facet(z)$, the facet
of $P$ pierced by the ray $\{tz :\; t \ge 0\}$.
In the shadow-vertex simplex method, we assume that \phaseI provides us with
an initial objective vector $z_0$ and the initial $\facet(z_0)$.
Then \phaseII of the simplex method computes $\facet(q)$ for all
vectors $q$ in the plane $E = \Span(z_0,z)$ between $z_0$ and $z$.
Specifically, it rotates $q$ from $z_0$ toward $z$ and
updates $\facet(q)$ by removing and adding one vertex to
its basis, as it becomes necessary. At the end, it outputs $\facet(z)$.

The number of pivot steps in the simplex method is bounded by the number
of facets of $P$ the plane $E$ intersects. This is the {\em size of the
planar section of the random polytope} $P$, the number of the edges
of the polygon $P \cap E$. Under a hypothetical assumption that $E$ is
fixed or is statistically independent of $P$, the average size of $P \cap E$
is polynomial in the dimension $d$ and the reciprocal of the standard deviation $\s$
and polylogarithmic in the number of vertices $n$, see Theorem~\ref{main section}
below.

The main complication of the analysis in \cite{ST} was that plane
$E = \Span(z_0,z)$ was also random, and moreover correlated
with the random polytope $P$.
It is not clear how to find the initial vector $z_0$ independent
of the polytope $P$ and, at the same time, in such a way that we know the
facet of $P$ it pierces. Thus the main problem rests in \phaseI.
None of the previously available \phaseI methods in linear programming seem
to effectively overcome this difficulty.

The randomized \phaseI proposed in \cite{ST} exposed a random facet of $P$
by multiplying a random $d$-subset of the vectors $a_i$ by an appropriately
big constant to ensure that these vectors do span a facet. Then a random
convex linear combination of these vectors formed the initial vector $z_0$.
This approach brings about two complications:

(a) the vertices of the new random polytope are no longer Gaussian;

(b) the initial objective vector $z_0$ (thus also the plane $E$)
  is correlated with the random polytope.

\noindent Our new approach will overcome both these difficulties.

\subsection{Phase-I for arbitrary linear programs}

We propose the following randomized \phaseI for arbitrary unit linear programs.
It is of independent interest, regardless of its applications to smoothed
analysis and to the simplex method.

Instead of finding or exposing a facet of $P$, we {\em add a facet} to $P$ in a random
direction. We need to ensure that this facet falls into the {\em numb set} of the linear
program, which consists of the points that do not change the solution when
added to the set of constraint vectors $(a_i)$.
Since the solution of the linear program is $\facet(z)$,
the affine half-space below the affine span of $\facet(z)$ (on the same side
as the origin) is contained in the numb set.
Thus {\em the numb set always contains a half-space}.

A random vector $z_0$ drawn from the uniform distribution on the
sphere $S^{d-1}$ is then in the numb half-space with probability
at least $1/2$. Moreover, a standard concentration of measure argument shows that
such a random point is at distance $\Omega(d^{-1/2})$ from
the boundary of the numb half-space, with constant probability.
(This distance is the {\em observable diameter} of the sphere, see \cite{L} Section~1.4). Thus a small regular simplex
with center $z_0$ is also in the numb set with constant probability.
Similarly, one can smooth the vertices of the simplex (make them Gaussian)
without leaving the numb set.
Finally, to ensure that such simplex will form a facet of the new
polytope, it suffices to dilate it by the factor
$M = \max_{i=1,\ldots,n} \|a_i\|$.

Summarizing, {\em we can add $d$ linear constraints to any linear program at random,
without changing its solution with constant probability}.

Note that it is easy to check whether the solution is correct, i.e. that the added
constraints do not affect the solution.
The latter happens if and only if none of the added constraints turn into equalities
 on the solution $x$.
Therefore, one can repeatedly solve the linear program with different
sets of added constraints generated independently, until the solution is correct.
Because of a constant probability of success at every step,
this \phaseI terminates after an expected constant number of steps,
and it always produces a correct initial solution.

When applied for the smoothed analysis of the simplex method, this \phaseI
resolves one of the main difficulties of the approach in \cite{ST}.
The initial objective vector $z_0$ and thus the plane $E$ become
independent of the random polytope $P$.
Thus {\em the smoothed complexity of the simplex method gets bounded
by the number of edges of a planar section of a random polytope $P$},
whose vertices have standard deviation of order
$\Omega^*(\min(\s, d^{-3/2}))$, see \eqref{ellsigma}.
In the previous approach \cite{ST}, such reduction was made with
the standard deviation of order $\Omega^*( n^{-14} d^{-8.5} \s^5)$.

\medskip

A {\em deterministic \phaseI} is also possible, along the same lines.
We have used that a random point in $S^{d-1}$ is at distance
$\Omega(d^{-1/2})$ from a half-space. The same property is clearly
satisfied by at least one element of the set $\{\pm e_1, \ldots, \pm e_d\}$,
where $\{e_1,\ldots,e_d\}$ is the canonical basis of $\R^d$.
Therefore, at least one of $d$ regular simplices of radius
$\frac{1}{2} d^{-1/2}$ centered at points $e_i$, lies in the numb half-space.
One can try them all for added constraints; at least one will
give a correct solution. This however will increase the running time by a
factor of $d$ -- the number of trials in this deterministic \phaseI may be
as large as $d$, while the expected number of trials in the
randomized \phaseI is constant. The smoothed analysis with such \phaseI will
also become more difficult due to having $d$ non-random vertices.

\subsection{Sections of randomly perturbed polytopes}  \label{s:remaining difficulties}

To complete the smoothed analysis of the simplex algorithm,
it remains to bound the size (i.e. the number of edges)
of the section $P \cap E$ of a randomly perturbed polytope $P$
with a fixed plane $E$.

Spielman and Teng proved the first polynomial bound on this size.
We improve it to a polylogarithmic bound in the number of vertices $n$:

\begin{theorem}[Sections of random polytopes]       \label{main section}
  Let $a_1,\ldots,a_n$ be independent Gaussian vectors in $\R^d$
  with centers of norm at most $1$, and with standard deviation $\s$.
  Let $E$ be a plane in $\R^d$. Then the random polytope
  $P = \Conv(a_1,\ldots,a_n)$ satisfies
  $$
  \E \, |\edges (P \cap E)| = O(d^5 \log^2 n + d^3 \s^{-4}).
  $$
\end{theorem}

See Theorem~\ref{section} for a slightly more precise result.

Spielman and Teng obtained a weaker estimate $O(n d^3 \s^{-6})$ for this size
(\cite{ST} Theorem~4.0.1). A loss of the factor of $n$ in their argument occurs in
estimating the angle of incidence (\cite{ST} Lemma~4.2.1), the angle at which
a fixed ray in $E$ emitted from the origin meets the facet of $P$ it pierces.

Instead of estimating the angle of incidence from one viewpoint determined by the
origin $0$, we will {\em view the polytope $P$ from three different points}
$0_1$, $0_2$, $0_3$ on $E$.
Rays will be emitted from each of these points, and from at least one
of them the angle of incidence will be good
(more precisely, the angle to the edge of $P \cap E$, which is the intersection
of the corresponding facet with $E$).

\section{Preliminaries}

\subsection{Notation}                   \label{s:notation}

The {\em direction} of a vector $x$ in a vector space is the ray
$\{tx :\; t \ge 0\}$.
The non-negative {\em cone} of a set $K$, denoted $\cone(K)$,
is the union of the directions of all the vectors $x$ in $K$.
The closed {\em convex hull} of $K$ is denoted by $\Conv(K)$ and $\conv(K)$.
(We prefer the latter notation when $K$ consists of $d$ points in $\R^d$).

The standard inner product in $\R^d$ is denoted by $\< x,y \> $,
and the standard Euclidean norm in $\R^d$ is denoted by $\|x\|$.
The unit Euclidean sphere in $\R^d$ is denoted by $S^{d-1}$.

The {\em polar} of a set $K$ in $\R^d$ is defined as
$K^\circ = \{ x \in \R^d : \; \< x,y \> \le 1 \; \forall y \in K \}$.

A {\em half-space} in $\R^d$ is a set of the form $\{x :\; \< z, x\> \le 0\}$ for
some vector $z$. An {\em affine half-space} takes the form
$\{x :\; \< z, x\> \le a\}$ for some vector $z$ and a number $a$.
The definitions of a {\em hyperplane} and {\em affine hyperplane} are similar,
with equalities in place of the inequalities.
The {\em normal to an affine hyperplane} $H$
that is not a hyperplane is the vector $h$ such that
$H = \{x :\; \< h, x\> = 1\}$.
A point $x$ is said to be {\em below} $H$ if $\< h,x\> \le 1$.
We say that a direction $z$ {\em pierces} the hyperplane $H$
(or a subset $H_0$ thereof) if the ray $\{tz :\; t \ge 0\}$ intersects $H$
(respectively, $H_0$).

The probability will be denoted by $\P$, and the expectation by $\E$.
The {\em conditional probability} on a measurable subset $B$ of a probability
space is denoted by $\P\{ \cdot | B\}$ and  defined as
$\P\{A|B\} = \P\{A \cap B\} / \P\{B\}$.

A {\em Gaussian random vector} $g = (g_1,\ldots,g_d)$ in $\R^d$ with
center $\bar{g} = (\bar{g}_1,\ldots,\bar{g}_d)$ and variance $\s$
is a vector whose coordinates $g_i$ are independent Gaussian random
variables with centers $\bar{g}_i$ and variance $\s$.

Throughout the paper, we will assume that the vectors $(a_i, b_i)$ that
define the linear program (LP) are in general position.
This assumption simplifies our analysis and it holds with probability
$1$ for a smoothed program. One can remove this assumption with
appropriate modifications of the results.

A solution $x$ of (LP) is determined by a $d$-set $I$ of the indices of
the constraints $\< a_i, x \> \le b_i$ that turn into equations on $x$.
One can compute $x$ from $I$ by solving these equations. So we
sometimes call the index set $I$ a solution of (LP).

For a polytope $P = \Conv(0,a_1,\ldots,a_n)$ and a vector $z$ in $\R^d$,
$\facet(z) = \facet_P(z)$ will denote the set of all faces of $P$
which the direction $z$ pierces. The point where it pierces them
(which is unique) is denoted by $z_P$.
More precisely, $\facet(z)$ is the family of all
$d$-sets $I$ such that $\conv(a_i)_{i \in I}$
is a facet of the polytope $P$ and the direction $z$ pierces it.

If $z$ is in general position, $\facet(z)$  is an
empty set or contains exactly one set $I$.
The corresponding geometric facet $\conv(a_i)_{i \in I}$ will be
denoted by $\Facet_P(z)$ or $\Facet_P(I)$.

Positive absolute constants will be denoted by $C,C_1,c,c_1,\ldots$.
The natural logarithms will be denoted by $\log$.

\subsection{Vertices at infinity}           \label{s:at infinity}
For convenience in describing the interpolation method,
we will assume that one of the constraint vectors $a_i$ can be {\em at infinity},
in a specified direction $u \in \R^d$.
The definitions of the positive cone and the convex hull are
then modified in a straightforward way.
If, say, $a_j$ is such an infinite vector and $j \in I$, then one defines
$\conv(a_i)_{i \in I} = \conv(a_i)_{i \in I-\{j\}} + \{tu : \; t \ge 0\}$,
where the addition is the Minkowski sum of two sets,
$A+B = \{a+b :\; a \in A, \; b \in B\}$.

Although having infinite vectors is convenient in theory,
all computations can be performed with numbers bounded by
the magnitude of the input (e.g., checking $I \in \facet(z)$
for given $z$ and $I$ has the same complexity whether or not some vertex of $P$
is at infinity).

\subsection{Polar shadow vertex simplex method} \label{s:PSVSM}
This is the only variant of the simplex method whose smoothed complexity
has been analyzed. We shall describe this method now; for more information
see \cite{ST} Section 3.2.

The polar shadow vertex simplex method works for unit linear programs,
i.e. programs (LP) with $b = \one$.
A solution of such program is a member of $\facet(z)$ of the polytope
$P = \Conv(0,a_1,\ldots,a_n)$. The program is unbounded iff
$\facet(z) = \emptyset$. (See \cite{ST} Section 3.2).

The input of the polar shadow vertex simplex method is:

\begin{enumerate}
  \item objective vector $z$;
  \item initial objective vector $z_0$;
  \item $\facet(z_0)$, provided that it consists of only one set of indices.
\end{enumerate}

The simplex method rotates $z_0$ toward $z$ and computes
$\facet(q)$ for all vectors $q$ between $z_0$ and $z$.
At the end, it outputs the limit of $\facet(q)$ as $q$ approaches $z$.
This is the last running $\facet(q)$ before $q$ reaches $z$.

If $\facet(z_0)$ contains more than one index set,
one can use the limit of $\facet(q)$ as $q$ approaches $z_0$
as the input of the simplex method. This will be the first
running $\facet(q)$ when $q$ departs from $z_0$.

If $z$ and $z_0$ are linearly dependent, $z_0 = -c z$ for some $c>0$,
one can specify an arbitrary {\em direction of rotation} $u \in \R^n$,
which is linearly
independent of $z$, so that the simplex method rotates $q$ in $\Span(z,u)$
in the direction of $u$, i.e. one can always write $q = c_1 z + c_2 u$ with
$c_2 \ge 0$.

\section{Reduction to unit programs: Interpolation}

We will show how to reduce an arbitrary linear program (LP) to a
{\em  unit linear program}
\begin{equation*}                       \tag{Unit LP}
\begin{aligned}
  &\text{maximize } \< z, x \>  \\
  &\text{subject to } Ax \le \one.
\end{aligned}
\end{equation*}
This reduction, whose idea originates from \cite{ST}, is quite general
and is independent from any particular method to solve linear programs.

The idea is to interpolate between (\UnitLP) and (LP).
To this end, we introduce an additional (interpolation) variable $t$
and a multiplier $\l$, and consider the {\em interpolated linear program}
with variables $x$, $t$:
\begin{equation*}                       \tag{\IntLP}
\begin{aligned}
  &\text{maximize } \< z, x \> + \l t \\
  &\text{subject to } Ax \le tb + (1-t)\one, \ \ \
                      0 \le t \le 1.
\end{aligned}
\end{equation*}
The interpolated linear program becomes (\UnitLP) for $t=0$
and (LP) for $t=1$. We can give bias to $t=0$ by choosing the multiplier
$\l \to -\infty$ and to $t=1$ by choosing $\l \to +\infty$.
Furthermore, (\IntLP) can be written as a unit linear program in $\R^{d+1}$:
\begin{equation*}                       \tag{\IntLP'}
\begin{aligned}
  &\text{maximize } \< (z, \l), (x,t) \>  \\
  &\text{subject to }
    \begin{cases}
      \< (a_i, 1-b_i), (x,t) \> \le 1, \\
      \< (0,1), (x,t) \> \le 1,\ \
      \< (0,-\infty), (x,t) \> \le 1.
    \end{cases}
\end{aligned}
\end{equation*}
The constraint vectors are $(a_i, 1-b_i)$, $(0,1)$ and $(0,-\infty)$.
(see Section~\ref{s:at infinity} about vertices at infinity).
This has a very useful consequence: if the constraints of the original (LP)
are Gaussian, then so are the constraints of (\IntLP'), except the two
last ones. In other words, {\em the reduction to a unit program preserves
the Gaussian distribution of the constraints}.

The properties of interpolation are summarized in the following
elementary fact.

\begin{proposition}[Interpolation]\hfill                \label{interpolation}

  (i) {\bf (LP) is unbounded} iff (\UnitLP) is unbounded
      iff (\IntLP) is unbounded for all sufficiently big $\l$
      iff (\IntLP) is unbounded for some $\l$.

  (ii) Assume (LP) is not unbounded.
      Then the solution of (\UnitLP) equals the solution of (\IntLP)
      for all sufficiently small $\l$; in this solution, $t=0$.

  (iii) Assume (LP) is not unbounded.
      Then {\bf (LP) is feasible} iff $t=1$ in the solution of (\IntLP)
      for all sufficiently big $\l$.

  (iv) Assume (LP) is feasible and bounded.
      Then the {\bf solution of (LP)} equals the solution of (\IntLP)
      for all sufficiently big $\l$.
\end{proposition}

\proof See Appendix \ref{a:interpolation}. \endproof

Now assuming that we know how to solve unit linear programs,
we will be able to solve arbitrary linear programs.
The correctness of this two-phase algorithm follows immediately
from Proposition~\ref{interpolation}.

\bigskip

\textsc{Solver for (LP)}

\nopagebreak

\fbox{\parbox{\algorithmwidth}{
  \begin{description}
    \item[Phase-I]
     Solve (\UnitLP) using \textsc{Solver for (Unit~LP)} of Section~\ref{s:phase-I}.
     If this program is unbounded, then (LP) is also unbounded.
     Otherwise, the solution of (\UnitLP) and $t=0$ is a limit solution of
     (Int LP) as $\l \to -\infty$.
     Use this solution as the input for the next step.

    \item[Phase-II]
     Use the polar shadow-vertex simplex method to find a limit solution
     of (\IntLP) with $\l \to +\infty$.
     If $t \ne 1$ in this solution, then the (LP) is infeasible.
     Otherwise, this is a correct solution of (LP).
  \end{description}
 }}

\bigskip

While this algorithm is stated in terms of limit solutions, one does not
need to take actual limits when computing them. This follows from the properties
of the polar shadow-vertex simplex method described in Section~\ref{s:PSVSM}.
Indeed, in \phaseII of \textsc{Solver for (LP)} we can write (\IntLP) as (\IntLP')
and use the initial objective vector $\bar{z}_0 = (0,-1)$,
the actual objective vector $\bar{z} = (0,1)$,
and the direction of rotation $\bar{u} = (z,0)$.
Phase-I provides us with a limit solution for the objective vectors
$(\e z, -1) = \bar{z}_0 + \e \bar{u}$ as $\e \to 0^+$.
These vectors approach $z_0$ as we rotate from $z$ toward $z_0$ in $\Span(z,u)$.
Similarly, we are looking for a limit solution for the objective
vectors $(\e z, 1) = \bar{z} + \e \bar{u}$ as $\e \to 0^+$.
These vectors approach $z$ as we rotate from $z_0$ toward $z$ in $\Span(z,u)$.
By Section~\ref{s:PSVSM}, the polar shadow-vertex simplex method
applied with vectors $\bar{z}_0$, $\bar{z}$, $\bar{u}$ and the initial limit solution
found in Phase-I, finds the correct limit solution in Phase-II.

\section{Solving unit programs: Adding constraints in Phase-I}               \label{s:phase-I}

We describe a randomized \phaseI for solving arbitrary unit linear problems
of type (Unit~LP). Rather than finding an initial feasible vertex,
we shall add a random vertex to the feasible set.
We thus add $d$ constraints to (Unit~LP), forming
\begin{equation*}                       \tag{Unit LP$^+$}
\begin{aligned}
  &\text{maximize } \< z, x \>  \\
  &\text{subject to } A^+ x \le \one,
\end{aligned}
\end{equation*}
where $A^+$ has the rows $a_1, \ldots, a_n, a_{n+1}, \ldots, a_{n+d}$
with some new constraint vectors $a_{n+1}, \ldots, a_{n+d}$.

The first big question is whether the problems (Unit~LP) and (Unit~LP$^+$)
are {\em equivalent}, i.e. whether
(Unit~LP$^+$) is bounded if and only if (Unit~LP) is bounded, and
if they are bounded, the solution of (Unit~LP$^+$) equals the solution
of (Unit~LP).
This motivates:

\begin{definition}
  A {\em numb set} of a unit linear program is the set of all vectors $a$
  so that adding the constraint $\< a, x\> \le 1$ to the set of the constraints
  produces an equivalent linear program.
\end{definition}

We make two crucial observations -- that the numb set is always big,
and that one can always check if the problems (Unit~LP) and (Unit~LP$^+$)
are equivalent. As mentioned in Section~\ref{s:notation}, we will assume
that the constraint vectors $a_i$ are in general position.

\begin{proposition}                     \label{numb set}
  The numb set of a unit linear program contains a half-space
  (called a {\em numb half-space}).
\end{proposition}

\begin{proof}
Given a convex set $K$ containing the origin in a vector space,
Minkowski functional $\|z\|_K$ is defined for vectors $z$ as
$\|z\|_K = \inf\{ \l > 0 :\; \frac{1}{\l} z \in K \}$ if the infimum
exists, and infinity if it does not exist.
Then the duality shows that the solution $\max_{Ax \le \one} \< z, x\> $
of (Unit~LP) equals $\|z\|_P$. (It is infinity iff the problem is unbounded;
we will use the convention $1/\infty = 0$ in the sequel).
By Hahn-Banach (Separation) Theorem, there exists a vector $z^*$ such that
$$
\< z^*, x\> \le \< z^*, \textstyle{\frac{1}{\|z\|_P}} z \> := h
\ \ \text{for all $x \in P$}.
$$
$0 \in P$ implies that $h \ge 0$. We define the affine half-space
$$
H^- = \{x :\; \< z^*, x\>  \le h \}
$$
and claim that $H^-$ lies in the numb set of (Unit~LP).
To prove this, let $a \in H^-$. Since $P \subset H^-$, we have
$\Conv(P \cup a) \subset H^-$, thus
$$
\|z\|_P \ge \|z\|_{\Conv(P \cup a)} \ge \|z\|_{H^-} = \|z\|_P
$$
where the first two inequalities follow from the inclusion
$P \subset \Conv(P \cup a) \subset H^-$, and the last equality
follows from the definition of $H^-$.
So, we have shown that $\|z\|_{\Conv(P \cup a)} = \|z\|_P$,
which says that $a$ and thus the affine half-space $H^-$
is in the numb set of (Unit LP). Since $h \ge 0$, $H^-$ contains
the origin, thus contains a half-space.
\end{proof}

In particular, if (Unit~LP) is bounded, then its numb set
is the affine half-space below $\facet(z)$.
Then a similar duality argument proves:

\begin{proposition}[Equivalence] \hfill             \label{equivalence}

  (i) If the added constraint vectors $a_{n+1},\ldots,a_{n+d}$ lie
    in some numb half-space of (Unit~LP),
    then (Unit~LP$^+$) is equivalent to (Unit~LP).

  (ii) (Unit~LP$^+$) is equivalent to (Unit~LP) if and only if
    either (Unit~LP$^+$) is unbounded or its solution does not
    satisfy any of the added constraints $\< a_i, x\> \le 1$,
    $i = n+1,\ldots,n+d$.  \endproof
\end{proposition}

Proposition~\ref{numb set} implies that a constraint vector $z_0$ whose
direction is chosen at random in the unit sphere $S^{d-1}$, is in the
numb set with probability at least $1/2$. By a standard concentration
of measure argument, a similar statement will be true about a small simplex
centered at $z_0$. It is then natural to take the vertices of this simplex
as added constraint vectors $a_{n+1}, \ldots, a_{n+d}$ for (Unit~LP$^+$).
To this end, we define the size $\ell$ of the simplex and the standard deviation $\s_1$
for smoothing its vertices as
\begin{equation}                        \label{ellsigma}
\ell = \frac{c_1}{\sqrt{\log d}}, \ \ \
\s_1 = \min \Big( \frac{1}{6 \sqrt{d \log n}}, \; \frac{c_1}{d^{3/2} \log d} \Big),
\end{equation}
where $c_1 = \frac{1}{300}$ and $c_2 = \frac{c_1^2}{100}$.
Then we form (Unit~LP$^+$) as follows:

\bigskip

\textsc{Adding Constraints}

\nopagebreak

\fbox{\parbox{\algorithmwidth}{
  \begin{description}
    \item[Input] Size $M_0>0$ and rotation $U \in O(d)$.
    \item[Output] ``Failure''
            or vectors $a_{n+1},\ldots,a_{n+d}$ and
                $z_0 \in \cone(a_{n+1}, \ldots, a_{n+d})$.
  \end{description}
  \begin{enumerate}
    \item {\em Form a regular simplex:}
     let $z'_0$ be a fixed unit vector in $\R^d$
     and $\bar{a}'_{n+1}, \ldots, \bar{a}'_{n+d}$
     be the vertices of a fixed regular simplex
     in $\R^d$ with center and normal $z'_0$,
     and radius $\|z'_0 - \bar{a}'_i\| = \ell$.

    \item {\em Rotate and dilate:}
     let $z_0 = 2M_0 Uz'_0$, \
     $\bar{a}_i = 2M_0 U \bar{a}'_i$ for $i=n+1,\ldots,n+d$.

    \item {\em Smooth:}
     let $a_i$ be independent Gaussian random variables with mean
     $\bar{a}_i$ and standard deviation $2M_0 \s_1$, for $i=n+1,\ldots,n+d$.

    \item {\em Check if the constraints added correctly:} check if
      \begin{enumerate}
        \item $z_0 \in \cone(a_{n+1},\ldots,a_{n+d})$ and
        \item distance from $0$ to $\aff(a_{n+1},\ldots,a_{n+d})$
         is at least $M$.
      \end{enumerate}
     If not, return ``Failure''.
  \end{enumerate}
 }}

\bigskip

\begin{remark}
  Steps (3) and (4) are, strictly speaking, not necessary.
  They facilitate the theoretical smoothed analysis
  of the simplex method. However, they can be skipped in
  practical implementations.
\end{remark}

\medskip

The crucial property of \textsc{Adding Constraints} is the following.
(Recall that we regard a solution of a linear program as the index set
of the inequalities that become equalities on the solution point,
see Section~\ref{s:notation}).

\begin{theorem}                         \label{adding constraints}
  Let (Unit~LP) be a unit linear program with a numb half-space $H$,
  and let $M_0 \ge M$ where $M = \max_{i=1,\ldots,n} \|a_i\|$.
  Then:

  1. Let $U \in O(d)$ be arbitrary. If the algorithm \textsc{Adding Constraints}
    does not return ``Failure'', then a solution of (Unit~LP$^+$) with the objective
    function $\< z_0, x\> $ is the index set $\{n+1,\ldots,n+d\}$.

  2. With probability at least $1/4$ in the choice of a random rotation $U \in O(d)$
    and random vectors $a_{n+1},\ldots,a_{n+d}$,
    the algorithm \textsc{Adding Constraints}
    does not return ``Failure'' and the vectors $a_{n+1},\ldots,a_{n+d}$
    lie in the numb half-space $H$.
\end{theorem}

\proof See Appendix \ref{a:adding constraints}. \endproof

\noindent By Proposition~\ref{equivalence},
the conclusion of Theorem~\ref{adding constraints} is that:

  (a) with constant probability the problems (Unit~LP$^+$) and (Unit~LP) are
    equivalent;

  (b) we can check whether they are equivalent or not (by part (ii) of
    Proposition~\ref{equivalence});

  (c) we always know a solution of (Unit~LP$^+$) for some objective function
    (if ``Failure'' is not returned).

\noindent Thus we can solve (Unit~LP) by repeatedly solving (Unit~LP$^+$)
with independently added constraints until no ``Failure'' is returned
and until the solution is correct.
This forms a two-phase solver for unit linear programs.

\bigskip

\textsc{Solver for (Unit~LP)}

\nopagebreak

\fbox{\parbox{\algorithmwidth}{
  Do the following until no ``Failure'' is returned and the solution $I^+$
  contains none of the indices $n+1,\ldots,n+d$:

  \begin{description}
    \item[Phase-I]
      Apply \textsc{Adding Constraints} with $M_0 = e^{\lceil \log M \rceil}$
      where $M = \max_{i=1,\ldots,n} \|a_i\|$
      and with the rotation $U$ chosen randomly and independently in the orthogonal
      group $O(d)$ according to the Haar measure.
      If no ``Failure'' returned, then $\{n+1,\ldots,n+d\}$ is a solution of
      (Unit~LP$^+$) with the objective function $\< z_0, x\> $.
      Use this solution as the input for the next step.

    \item[Phase-II]
      Use the polar shadow-vertex simplex method to find a solution $I^+$ of
      (Unit~LP$^+$) with the actual objective function $\< z, x\> $.
  \end{description}
 Return $I^+$.
 }}

\bigskip

\begin{remark}
  The discretized maximum $M_0$ is introduced in this algorithm only
  to simplify its smoothed analysis.
  In practical implementations of the algorithm, one can use $M_0 = M$.
\end{remark}

\section{Bounding the complexity via sections of random polytopes} \label{s:pivots}

We do here the smoothed analysis of \textsc{Solver for (LP)}, and
prove the following more precise version of Theorem~\ref{main}:

\begin{theorem}                        \label{main precise}
  For an arbitrary linear program with $d>3$ variables and $n>d$ constraints,
  the expected number of pivot steps
  in a two-phase shadow-vertex simplex method for the smoothed program is
  $$
  O \big( \log^2 n \cdot \log\log n \cdot
  (d^3 \s^{-4} + d^5 \log^2 n + d^9 \log^4 d) \big)
  = O^*(d^9 + d^3 \s^{-4}).
  $$
\end{theorem}

\begin{remark}
  A more careful analysis may allow one to remove the factor
  $\log^2 n \cdot \log \log n$.
  To do this, one uses the version of the algorithm with $M_0 = M$
  and bounds the sections of a polytope with an added facet.
  This makes the analysis a bit harder, because the magnitudes
  of the added vertices correlate with $M$ and thus with the
  polytope.
\end{remark}

\medskip

To prove Theorem~\ref{main precise}, let us recall how many calls
\textsc{Solver for (LP)} makes to the
polar shadow-vertex method. One call is made to solve (\IntLP)
(in the second phase), and several calls are made to solve (\UnitLP$^+$)
(in the subroutine \textsc{Solver for (Unit~LP)} in the first phase.)

The expected number of calls (iterations) in \textsc{Solver for (Unit~LP)}
is $4$. This follows from part~2 of Theorem~\ref{adding constraints}
and Proposition~\ref{equivalence}. Thus:

\begin{quote}{\em
  The running time of
  \textsc{Solver for (LP)} is bounded by the total number of
  pivot steps made in the polar shadow-vertex simplex method,
  when we apply it:
  \begin{enumerate}
    \item once for (Int~LP);
    \item on average, four times for (Unit~LP$^+$).
  \end{enumerate} }
\end{quote}

Furthermore, as explained in Section~\ref{s:duality},
the number of pivot steps
in the polar shadow-vertex simplex method on a unit linear
program is bounded by the {\em number of edges of the polygon $P \cap E$},
where $P$ is the convex hull of the origin and the constraint vectors,
and $E$ is the span of the initial and the actual objective vectors.

\medskip

We shall now first estimate the size of the section of the polytope
for (1) and (2) separately, and then combine them by a simple
stopping time argument.

Recall that the vectors $a_i$ and $b$ are Gaussian random vectors with
centers $\bar{a}_i$ and $\bar{b}$ respectively, and with standard
deviation $\s$. We can assume without loss of generality that
\begin{equation}                        \label{ai bi sigma}
\|(\bar{a}_i, \bar{b}_i)\| \le 1
\ \ \text{for all $i=1,\ldots,n$}, \ \ \ \
\s \le \frac{1}{6 \sqrt{d \log n}}.
\end{equation}
(To achieve these bounds, we scale down the vectors if necessary --
first to achieve $\max_i \|(a_i, b_i)\| = 1$, then further to make $\s$
as required).

\subsection{Sections of random polytopes}

When we apply the polar shadow-vertex simplex method for (Int~LP)
(in \phaseII of \textsc{Solver for (LP)}),
the plane
$$
E = \Span((z,0),(0,1))
$$
is fixed, and the constraint vectors are
$$
(0,1), \ (0,-\infty), \ \text{and } (a_i, 1-b_i)_{i=1}^n.
$$

The vertices $(0,1)$, $(0,-\infty)$ and the origin can be removed
from the definition of $P$ using the elementary observation that
if $a \in E$ then the number of edges of $\Conv(P \cup a) \cap E$
is at most the number of edges of $P \cap E$ plus $2$.
Since $(0,1)$, $(0,-\infty)$ and $0$ do lie in $E$, they can be ignored
at the cost of increasing the number of edges by $6$.
Thus we can assume that
$$
P = \Conv(a_i, 1-b_i)_{i=1}^n,
$$
where the vectors $(a_i, 1-b_i)$ are independent Gaussian vectors with centers
of norm at most $2$, and with standard deviation $\s$.

Scaling these vectors down so that their norms become at most $1$,
we deduce the desired size of the section $P \cap E$ from the
following theorem. It gives a desired bound for the number of
pivots when we solve (\IntLP).

\begin{theorem}[Sections of random polytopes]               \label{section}
  Let $a_1,\ldots,a_n$ be independent Gaussian vectors in $\R^d$
  with centers of norm at most $1$, and whose standard deviation $\s$ satisfies
  \eqref{ai bi sigma}.
  Let $E$ be a plane in $\R^d$. Then the random polytope
  $P = \Conv(a_1,\ldots,a_n)$ satisfies
  \begin{equation}                                  \label{section bound}
    \E \, |\edges (P \cap E)| \le C d^3 \s^{-4},
  \end{equation}
  where $C$ is an absolute constant.
\end{theorem}

\proof See Section~\ref{s:section}. \endproof

\begin{remark}
  Spielman and Teng obtained a weaker estimate $O(n d^3 \s^{-6})$ for this size
  (\cite{ST} Theorem~4.0.1). Because of the polynomial,
  rather than a polylogarithmic, dependence on $n$, their bound
  is not sufficient for us.
\end{remark}

\medskip

Summarizing, we have shown that:
\begin{equation}                    \label{IntLP pivots}
  \text{{\em (\IntLP) makes in expectation $D(d,\s) + 6$ pivot steps,}}
\end{equation}

where $D(d,\s)$ denotes the right hand side in \eqref{section bound}.

\subsection{Sections of a random polytope with an added facet}

When we repeatedly apply the polar shadow-vertex simplex method for (Unit~LP$^+$)
(in \textsc{Solver for (Unit~LP)}), each time we do so with $U$ chosen
randomly and independently of everything else. Let us condition on a choice
of $U$. Then the plane $E$ is fixed:
$$
E = \Span(\frac{z_0}{\|z_0\|}, z) = \Span(Uz'_0, z).
$$

The constraint vectors are
$$
a_1, \ldots, a_{n}, \; a_{n+1}, \ldots, a_{n+d}.
$$
The first $n$ of these are independent Gaussian vectors
with centers of norm at most $1$ and whose standard deviation $\s$
satisfies \eqref{ai bi sigma}.

The last $d$ are also Gaussian vectors chosen independently
with centers $2M_0\tilde{a}_i$ and variance $2M_0\s_1$, where
$\tilde{a}_i (= U \bar{a}'_i)$ are fixed vectors
of norm
$$
\|\tilde{a}\| = \|\bar{a}'_i\|
= \sqrt{ \|z_0'\|^2 + \|z_0' - \bar{a}'_i\|^2 }
= \sqrt{1+\ell^2} \le 1.01.
$$
(Here we used the orthogonality of $z_0'$ and $z_0' - a_i'$, which holds
by the construction of these vectors in algorithm \textsc{Adding Constraints}).

Recall that $M_0 = e^{\lceil \log M \rceil}$, where
$M = \max_{i=1,\ldots,n} \|a_i\|$
and $\s_1$ is as in \eqref{ellsigma}.

Let $\Phi(a_1,\ldots,a_{n+d})$ denote the density of such vectors
as above.
One should note that the last $d$ of vectors correlate with the
first $n$ vectors through the random variable $M_0$.
This difficulty will be resolved by an argument similar to that in \cite{ST}.
We will show that with high probability,
$M_0$ takes values in a small finite set.
For each fixed $M_0$ in this set, all the vectors $a_1,\ldots,a_{n+d}$
are independent, so we can use Theorem~\ref{section} to get the
desired size of the section $P \cap E$:

\begin{corollary}[Sections of random polytopes with an added facet]
                                                    \label{section+}
  Let $a_1,\ldots,a_{n+d}$ be random vectors in $\R^d$ with joint
  distribution $\Phi(a_1,\ldots,a_{n+d})$.
  Let $E$ be a plane in $\R^d$. Then the random polytope
  $P = \Conv(a_1,\ldots,a_{n+d})$ satisfies
  $$
  \E \, |\edges (P \cap E)|
    \le C \log \log n \cdot D(d,\s_0)
  $$
  where $D(d,\s)$ denotes the right hand side of \eqref{section bound},
  where
  $$
  \s_0 = c \log^{-1/2}n \cdot \min(\s,\s_1),
  $$
  and where $c$ is an absolute constant.
\end{corollary}

\proof See Section~\ref{s:section+}. \endproof

Thus, similarly to the previous section, we have shown that:
\begin{equation}                    \label{UnitLP pivots}
  \text{{\em (\UnitLP$^+$) makes in expectation
    $C \log \log n \cdot D(d,\s_0) + 6$ pivot steps}}.
\end{equation}

\subsection{The total number of pivot steps}

As we mentioned in the beginning of this section,
the total number of
\begin{equation}                            \label{grand total}
  \text{pivot steps in \textsc{Solver for (LP)}} = Y + Z,
\end{equation}
where $Y$ is the number of pivot steps to solve (\IntLP),
and $Z$ is the total number of pivot steps to solve (\UnitLP$^+$)
over all iterations \textsc{Solver for (Unit~LP)} makes.

Then \eqref{IntLP pivots} states that
\begin{equation}                            \label{EY}
 \E Y \le D(d,\s) + 6.
\end{equation}
Furthermore, the expected number of iterations in
\textsc{Solver for (Unit~LP)} is at most four.
Then \eqref{UnitLP pivots} yields a good bound for $Z$.
This is rigorously proved by the following simple stopping time argument.

Consider a variant of \textsc{Solver for (Unit~LP)}, from
which the stopping condition is removed, i. e. which repeatedly
applies \phaseI and \phaseII in an infinite loop.
Let $Z_k$ denote the number of pivot steps in \phaseII
of this algorithm in $k$-th iteration, and $F_k$ denote the random
variable which is $1$ if $k$-th iteration in this algorithm results in
failure, and $0$ otherwise.
Then the expected total number of pivot steps made in the
actual \textsc{Solver for (Unit~LP)}, over all iterations, is
distributed identically with
$$
Z \equiv \sum_{k=1}^\infty Z_k \prod_{j=1}^{k-1} F_j.
$$
To bound the expectation of $Z$, we denote by $\E_0$ the expectation
with respect to random (smoothed) vectors $(a_1, \ldots, a_n)$,
and by $\E_j$ the expectation with respect to the random choice made
in $j$-th iteration of \textsc{Solver for (Unit~LP)}, i. e. the
choice of $U$ and of $(a_{n+1},\ldots,a_{n+d})$.

Let us first condition on the choice of $(a_1,\ldots,a_n)$.
This fixes the numb set, which makes each $F_j$ depend only on
the random choice made in $j$-th iteration, while $Z_k$ will only
depend on the random choice made in $k$-th iteration. Therefore
\begin{equation}                \label{EZ}
\E Z = \E_0 \sum_{k=1}^\infty (\E_k Z_k) \prod_{j=1}^{k-1} \E_j F_j.
\end{equation}
As observed above,
$$
\E_j F_j  = \P (F_j = 1) \le 3/4,
$$
which bounds the product in \eqref{EZ} by $(3/4)^k$.
Moreover, $\E_0 \E_k Z_k$ are equal to the same value for all $k$ because
of the identical distribution. Thus
$$
\E_0 \E_k Z_k \le \max_U \E_\Phi Z_1,
$$
where $E_\Phi$ is the expectation with respect to the random vectors
$(a_1,\ldots,a_{n+d})$ conditioned on a choice of $U$ in the first iteration.
Thus
$$
\E Z \le 4 \max_U \E_\Phi Z_1.
$$

The random vectors $(a_1,\ldots,a_{n+d})$ have joint density $\Phi$,
and we can apply \eqref{UnitLP pivots} to bound
$$
\max_U \E_\Phi Z_1 \le 2C \log \log n \cdot D(d,\s_0).
$$
Summarizing,
$$
\E Z \le 8 C \log \log n \cdot D(d,\s_0).
$$
This, \eqref{EY}, and \eqref{grand total} imply that
the expected number of pivot steps in \textsc{Solver for (LP)}
is at most
$$
D(d,\s) +6 + 8 C \log \log n \cdot D(d,\s_0)
\le C_1 \log^2 n \cdot \log\log n \cdot d^3 (\s^{-4} + d^2 \log^2 n + d^6 \log^4 d).
$$
This proves Theorem~\ref{main precise} and completes the smoothed analysis
of the simplex method.

\section{Sections of random polytopes with i.i.d. vertices} \label{s:section}

The goal of this section is to prove Theorem~\ref{section}
about the size of planar sections of random polytopes.

Our argument improves upon the part of the argument of \cite{ST}
where it looses a factor of $n$. Recall that we need a polylogarithmic
dependence on $n$.

In Section~\ref{s:counting}, we will outline the counting argument of \cite{ST},
the crucial idea of which is to reduce the counting problem to the geometric
problem of bounding a fixed point away from the boundary of a random simplex.
In Section~\ref{s:viewpoints}, we will give a simple ``three viewpoints''
argument that allows us to not loose a factor of $n$ in contrast to the
original argument of \cite{ST}.
In further sections, we revisit the counting argument and complete the proof.

\subsection{Spielman-Teng's counting argument}          \label{s:counting}

We start by outlining the counting argument of \cite{ST}.
This argument leads to a loss of a factor linear in $n$.
We shall show then how to improve this to a polylogarithmic factor.

We consider the one-dimensional torus in the plane $E$ defined as
$\T = E \cap S^{d-1}$. We parametrize it by
$$
q = q(\theta) = z \sin(\theta) + t \cos(\theta), \qquad \theta \in [0,2\pi),
$$
where $z$, $t$ are some fixed orthonormal vectors in $E$.

Assume for now that the origin always lies in the interior of the polytope $P$;
we will get rid of this minor assumption later. Then clearly
$$
\Exp := \E | \edges(P \cap E)| = \E |\{ \facet_P(q) :\; q \in \T \}|.
$$

In order to turn this into finite counting,
we quantize the torus $\T$ by considering
$$
\T_m := \{ \text{$m$ equispaced points in $\T$} \}.
$$
A simple discretization argument (see \cite{ST} Lemma~4.0.6) gives that
\begin{equation}                                    \label{E lim E}
  \E |\{ \facet_P(q) :\; q \in \T \}|
  = \lim_{m \to \infty} \E |\{ \facet_P(q) :\; q \in \T_m \}|
\end{equation}

\begin{center}
\raisebox{-1 true in}{\includegraphics[height=2in]{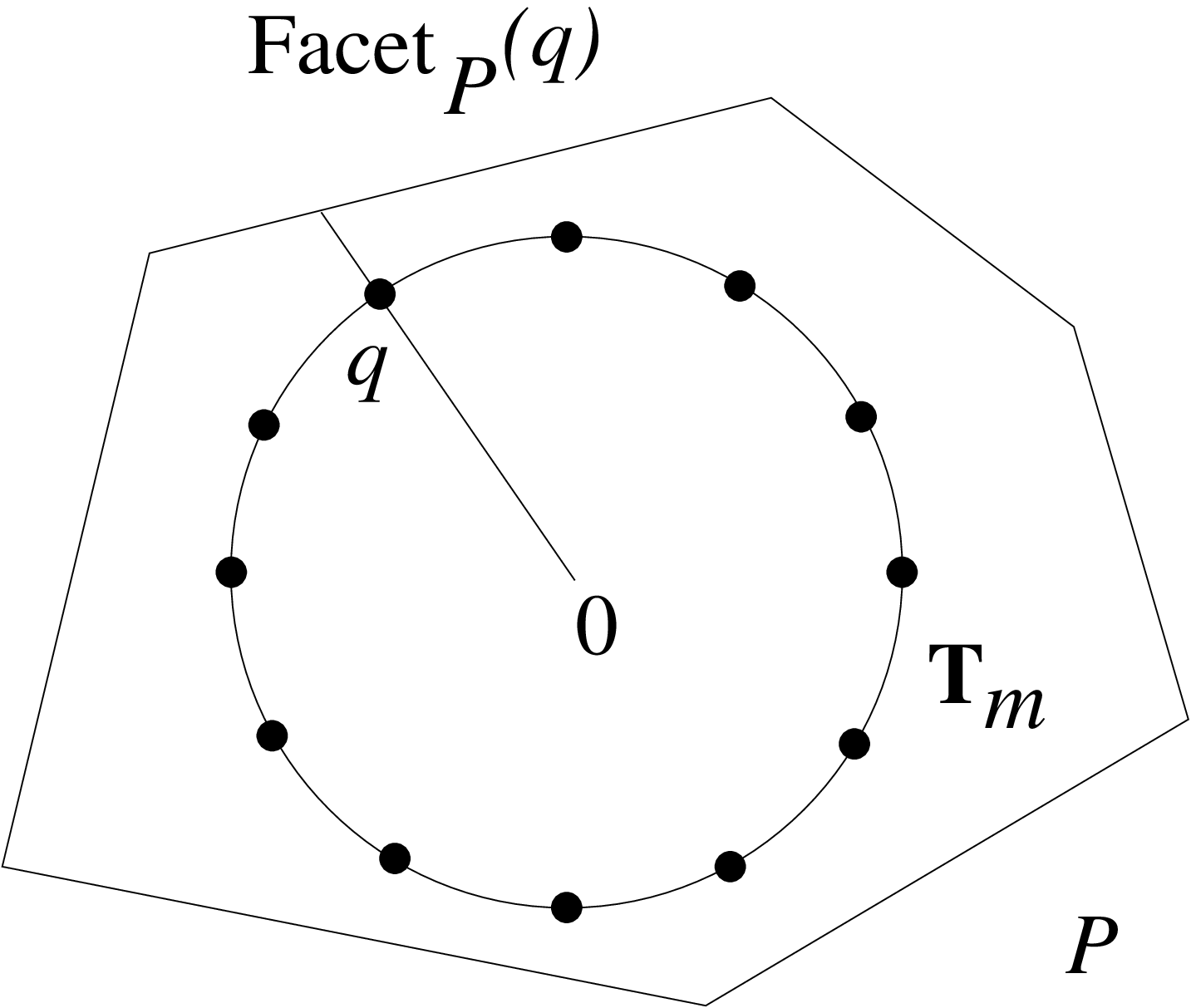}}
\end{center}

For the proper counting, one would prefer to keep one point $q$ per facet,
for example the one closest to the boundary of the facet.
The closeness here is measured with respect to the angular distance:

\begin{definition}[Angular distance]
  Let $x$ and $y$ be two vectors in $\R^d$. The {\em angular distance}
  $$
  \ang(x,y) := \cos^{-1} \Big( \frac{|\< x,y \> |}{\|x\| \|y\|} \Big)
  $$
  is the angle formed by the vectors $x$ and $y$.
  The {\em angular length} $\ang(\II)$ of an interval $\II$ in $\R^d$
  is the angular distance between its endpoints.
\end{definition}

So in the counting formula \eqref{E lim E}, we can leave
one $q$ per facet, namely the $q = q(\theta)$ with the maximal $\theta$ (according
to the parametrization of the torus in the beginning of the argument).
Therefore, the angular distance of such $q$ to the boundary of $\Facet_{P \cap E} (q)$
(one of the endpoints of this interval), and hence also to the boundary of
$\Facet_P(q)$, is at most $2\pi/m$.

We have proved that
$$
\Exp \le \lim_{m \to \infty} \E |\{ \facet_P(q) :\;
  \ang\big( q, \partial \Facet_P(q) \big) \le \frac{2\pi}{m}, \; q \in \T_m \}|.
$$
With a slight abuse of notation, we shall denote
the set of all $d$-subsets of $\{1,\ldots,n\}$ by $\binom{n}{d}$. Then
\begin{equation}                                \label{prelim sum sum}
\Exp \le \lim_{m \to \infty}
  \sum_{q \in \T_m} \sum_{I \in \binom{n}{d}}
  \P \{ \facet_P(q) = I \text{ and }
    \ang \big( q, \partial \conv(a_i)_{i \in I} \big) \le \frac{2\pi}{m} \}.
\end{equation}
For every $q$, there exists at most one $I \in \binom{n}{m}$ such that
$\facet_P(q) = I$. Hence
$$
\sum_{I \in \binom{n}{d}} \P \{ \facet_P(q) = I \} \le 1
$$
Using this bound in \eqref{prelim sum sum}
and estimating $\sum_{q \in \T_m}$ above by
$m \cdot \max_{q \in \T_m}$, we obtain
\begin{equation}                                \label{prelim Exp0}
  \Exp \le \lim_{m \to \infty} m \cdot p(m),
\end{equation}
where
$$
p(m) = \max_{q \in \T_m, I \in \binom{n}{d}}
  \P \{ \ang \big( q, \partial \conv(a_i)_{i \in I} \big) \le \frac{2\pi}{m}
    \; \big| \;
    \facet_P(q) = I \}.
$$

Furthermore, one can get rid of the polytope $P$ by analyzing the event $\facet_P(q) = I$.
We condition on the realization on the points $(a_i)_{i \in I^c}$, as well
as on the subspace $E_I := \Span(a_i)_{i \in I}$.
The randomness remains in the random points $a_i$, $i \in I$ inside the (fixed) subspace $E_I$.
Then the event $\facet_P(q) = I$
holds if and only if the (fixed) point $q_P$ where the direction of $q$ pierces $E_I$
lies in the the simplex $\conv(a_i)_{i \in I}$.

We have thus reduced the problem to {\em estimating the distance of a fixed point
to the boundary of a random simplex, conditioned on the event that the point lies
in that simplex}.

\medskip

The main difficulty is that the distance is angular rather than Euclidean;
the latter is easier to estimate.
Unfortunately, the two distances may be very different. This happens
if the {\em angle of incidence} -- the
angle at which the direction of $q$ meets the subspace $E_I$ -- is too small.
So Spielman and Teng needed to show that the angle of incidence
is at least of order $1/n$ with high probability (Section~4.2 in \cite{ST});
consequently, the angular and Euclidean distances are within a factor $O(n)$
from each other.

In this paper, we can not tolerate the loss of a factor of $n$ since we are proving
a complexity estimate that is polylogarithmic rather than polynomial in $n$.
We will now present a simple way to avoid such a loss.

\subsection{Three viewpoints}       \label{s:viewpoints}

Instead of estimating the angle of incidence from one viewpoint determined by the
origin $0$, we will {\em view the polytope $P_0$ from three different points}
$0_1$, $0_2$, $0_3$ on $E$.
Vectors $q$ will be emitted from each of these points, and from at least one
of them the angle of incidence will be good
(more precisely, the angle of $q$ to the intersection of its facet with $E$ will be good).
This is formalized in the following two elementary observations on the plane.

\begin{lemma}[Three viewpoints]                 \label{three viewpoints}
  Let $K = \Conv(b_1,\ldots,b_N)$ be a planar polygon,
  where points $b_i$ are in general position and have norms at most $1$.
  Let $0_1$, $0_2$, $0_3$ be the vertices of an equilateral triangle
  centered at the origin and with norm $4$. Denote $K_i = \Conv(0_i, K)$.
  Then, for every edge $(b_k,b_m)$ of $K$, there exists $i \in \{1,2,3\}$
  such that $(b_k,b_m)$ is an edge of $K_i$, and
  $$
  \dist(0_i, \aff(b_k,b_m)) \ge 1.
  $$
\end{lemma}

\begin{center}
\raisebox{-1 true in}{\includegraphics[height=1.8in]{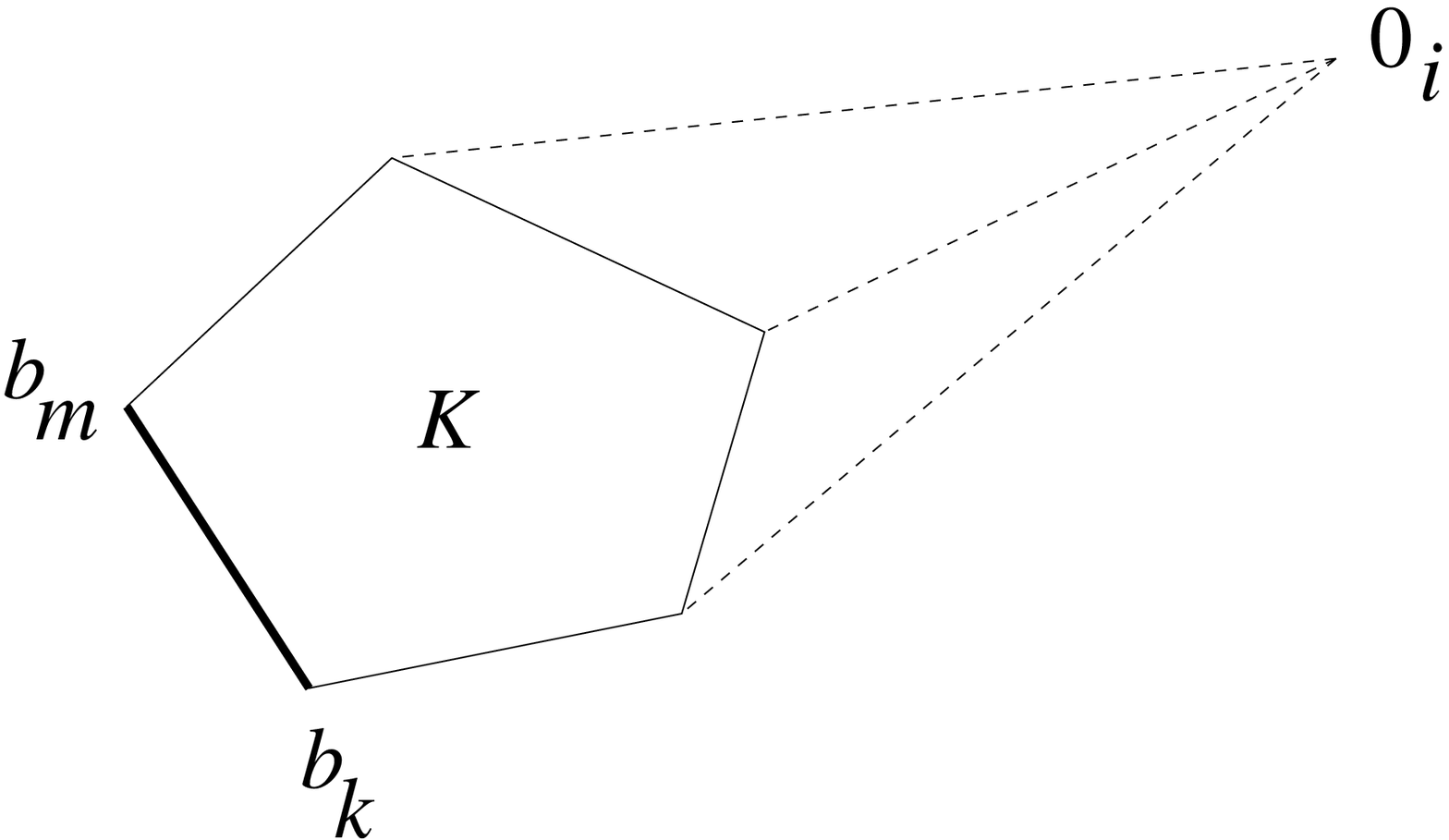}}
\end{center}

\begin{proof}
Let $L$ be any line passing through the origin. Then, for every equilateral
triangle centered at the origin and whose vertices have norms $R$,
there exist two vertices separated by the line $L$ and
whose distances to $L$ are at least $R/2$.
(The bound $R/2$ is attained if $L$ is parallel to one of the sides
of the triangle).

Let $L$ be the line passing through the origin and parallel to the
edge $(b_k,b_m)$.
It follows that among the three points $0_1$, $0_2$, $0_3$ there
exists at least two (say, $0_1$ and $0_2$) separated by the line $L$
and such that
$$
\dist(0_i, L) \ge 4/2 = 2, \qquad i=1,2.
$$
Moreover, since all the points $b_i$ have norms at most $1$, we have
$$
\dist(L, \aff(b_k,b_m)) \le \dist(0,b_k) \le 1.
$$
Then by the triangle inequality, $0_1$ and $0_2$ are separated by the
line $\aff(b_k,b_m)$ and
$$
\dist(0_i, \aff(b_k,b_m)) > 1, \qquad i=1,2.
$$

Since $0_1$ and $0_2$ are separated by the affine span of the
edge $(b_k,b_m)$ of the polygon $K$,
one of these points (say, $0_1$) lies on the same side from $\aff(b_k,b_m)$
as the polygon $K$. It follows that $(b_k,b_m)$ is an edge of $\Conv(0_1,K)$.
This completes the proof.
\end{proof}

\medskip

\begin{lemma}[Angular and Euclidean distances]          \label{angular euclidean}
  Let $L$ be a line in the plane such that
  $$
  \dist(0,L) \ge 1.
  $$
  Then, for every pair of points $x_1$, $x_2$ on $L$ of norm at most $10$,
  one has
  $$
  c \dist(x_1,x_2) \le \ang(x_1,x_2) \le \dist(x_1,x_2)
  $$
  where $c = (10^2+1)^{-1}$.
\end{lemma}

\begin{proof}
Without loss of generality, we may assume that $\dist(0,L) = 1$.
Choose unit vectors $u$ and $v$ in the plane such that $\< u, v \> = 0$
and $L = \{ u+tv :\; t \in \R\}$.
Then
$$
x_1 = u + t_1 v, \quad x_2 = u + t_2 v
$$
for some $t_1, t_2 \in \R$.

\begin{center}
\raisebox{-1 true in}{\includegraphics[height=1.5in]{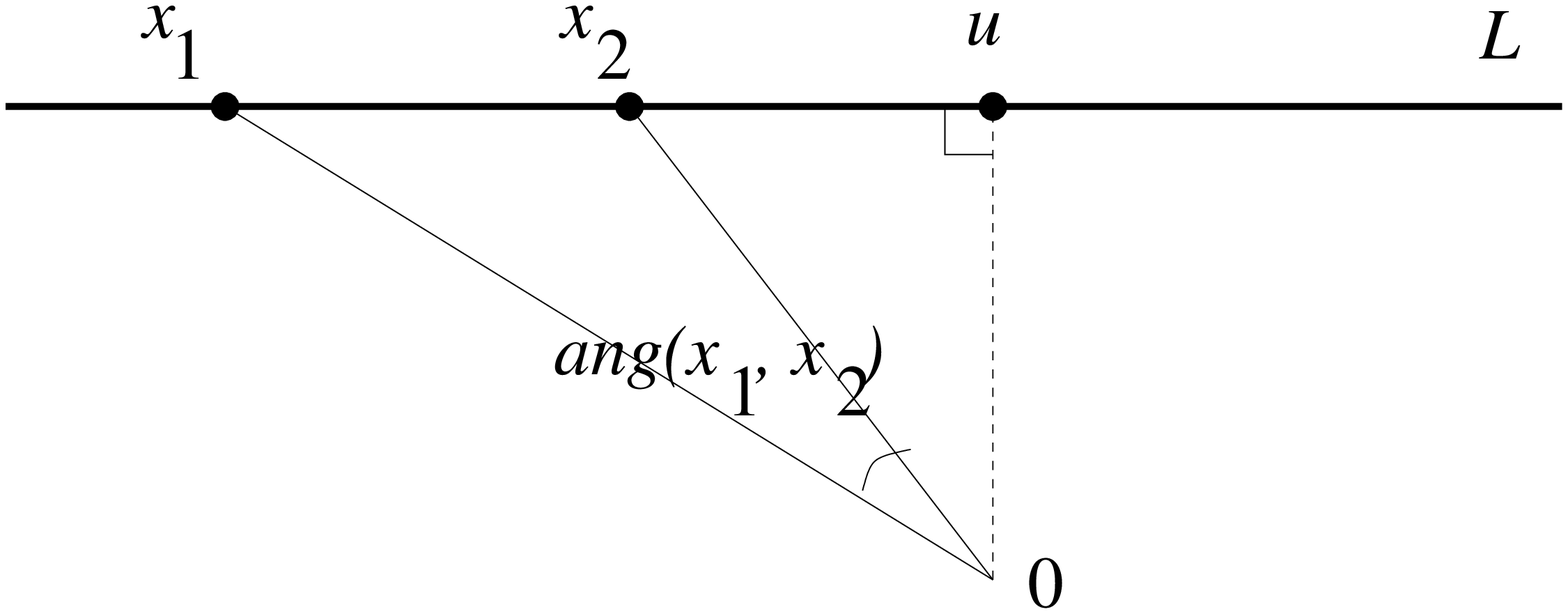}}
\end{center}

Hence we have
$$
|t_1| \le \|x_1\| \le 10, \quad
|t_2| \le \|x_2\| \le 10.
$$
Without loss of generality, $t_1 \le t_2$.
Then one easily checks that
$$
\dist(x_1,x_2) = t_2 - t_1, \qquad
\ang(x_1,x_2) = \tan^{-1} t_2 - \tan^{-1} t_1.
$$
Thus
$$
\ang(x_1,x_2) = \int_{t_1}^{t_2} \frac{d}{dt} (\tan^{-1} t) \; dt
= \int_{t_1}^{t_2} \frac{dt}{1+t^2}.
$$
We can estimate the integrand from both sides using the inequalities
$1 \le 1+t^2 \le 1+t_2^2 \le 1+10^2$.
Thus
$$
\frac{t_2-t_1}{1+10^2}
\le \ang(x_1,x_2)
\le t_2 - t_1.
$$
This completes the proof.
\end{proof}

\medskip

In view of Lemma~\ref{angular euclidean}, we can rephrase
Lemma~\ref{three viewpoints} as follows:
every edge (facet) of $K$ can be viewed from one of the
three viewpoints $0_1$, $0_2$, $0_3$ at a nontrivial angle,
and yet remain an edge of the corresponding polygon $\Conv(0_i,K)$.

\medskip

\subsection{Boundedness of the random polytope}

In Lemmas~\ref{three viewpoints} and \ref{angular euclidean}, the boundedness
requirements (for the points $x_i$ and $b_i$ respectively) are clearly
essential. To make sure that these requirements are satisfied in our setting,
we recall that the vertices of the polytope $P$ are i.i.d. Gaussian vectors.
We shall therefore use the following well-known estimate on the size of a Gaussian
vector, see e.g. Proposition 2.4.7 in \cite{ST}.

\begin{lemma}                           \label{gaussian}
  Let $g$ be a Gaussian vector in $\R^n$ ($n \ge 3$)
  with center $\bar{g}$ of norm at most $1$,
  and with variance $\s$. Then:

  1. We have $\P \{ \|g - \bar{g}\| \ge 3 \s \sqrt{d \log n} \}
     \le n^{-2.9 d}$.

  2. We have $\P \{ \|g\| \le c \s \sqrt{d} \} \le e^{-d}$,
    where $c = e^{-3/2}$.
\end{lemma}

We consider the event
$$
\EE := \{ \text{$\|a_i\| \le 2$, $i=1,\ldots,n$} \}.
$$
By Lemma~\ref{gaussian} and using our assumption \eqref{ai bi sigma},
we have
\begin{equation}                            \label{E}
  \P \{ \EE^c \}
  \le \P \{ \max_{i=1,\ldots,n} \|a_i\| > 3 \s \sqrt{d \log n} + 1 \}
  \le n^{-2.9d+1} \le 0.0015 \binom{n}{d}^{-1}.
\end{equation}

Our goal in Theorem~\ref{section} is to estimate
$$
\Exp := \E \, |\edges (P \cap E)|.
$$
The random variable $|\edges (P \cap E)|$ is bounded above
by $\binom{n}{d}$, which is the maximal number of facets of $P$.
Using \eqref{E}, it follows that
$$
\Exp := \E \, |\edges (P \cap E)|
\le \E \, |\edges (P \cap E)| \cdot \one_\EE + 1.
$$
We will use a similar intersection argument several times in the sequel.

\subsection{Counting argument revisited from three viewpoints}

We will apply Lemma~\ref{three viewpoints} in combination with \ref{angular euclidean}
for the random polygon $P \cap E$, whenever it satisfies $\EE$.
All of the points of this polygon are then bounded by $2$ in norm,
so we scale the result in Lemma~\ref{three viewpoints} by the factor $2$.

Let $0_1$, $0_2$, $0_3$ be the vertices of an equilateral triangle in
the plane $E$, centered at the origin and with
\begin{equation}                                \label{0i}
  \|0_1\| = \|0_2\| = \|0_3\| = 8.
\end{equation}
Denote
\begin{equation}                                \label{Pi}
  P_i = \Conv(0, -0_i + P).
\end{equation}

Lemma~\ref{three viewpoints} states in particular
that, if $\EE$ holds, then each edge of $P \cap E$
can be seen as an edge from one of the three
viewpoints. Precisely, there is a one-to-one
correspondence\footnote{
    Here and in the sequel, we identify $\facet(q)$ with the index set it contains.
    Since the polytope in question is almost surely in general position,
    $\facet(q)$ contains at most one index set.}
between the edges of $P \cap E$ and the set
$\{ \facet_{P_i \cap E}(q) : \; q \in \T, \; i=1,2,3 \}$.
We can further replace this set by
$\{ \facet_{P_i}(q) : \; q \in \T, \; i=1,2,3 \}$, since each $\facet_{P_i}(q)$
uniquely determines the edge $\facet_{P_i \cap E}(q)$; and vice versa,
each edge can belong to a unique facet.
Therefore
\begin{align}                           \label{discretize}
  \Exp &\le \E \, |\{ \facet_{P_i}(q) : \; q \in \T, \; i=1,2,3 \}| + 1 \\
  \intertext{and, by a discretization in limit as in Section~\ref{s:counting},}
  &= \lim_{m \to \infty} \E \, |\{ \facet_{P_i}(q) : \; q \in \T_m, \; i=1,2,3 \}| + 1.
\end{align}

Moreover, by the same discretization argument, we may ignore in \eqref{discretize}
all facets whose intersection with $E$ have angular length no bigger than, say,
$2\pi/m$.
After this, we replace $P_i$ by $P_i \cap E$ as we mentioned above,
and intersect with the event $\EE$ again, as before. This gives
\begin{equation}                            \label{exp in limit}
\Exp
\le  \lim_{m \to \infty} \E \,
  |\{ \facet_{P_i \cap E}(q) \ \text{of angular length} > \frac{2\pi}{m}; \;
  q \in \T_m, \; i=1,2,3 \}| \cdot \one_\EE + 2.
\end{equation}

\medskip

We are going to apply Lemmas~\ref{three viewpoints} and \ref{angular euclidean}
for a realization of $P$ for which the event $\EE$ holds.
Consider any facet from the set in \eqref{exp in limit}.
So let
$$
I = \facet_{P_i \cap E}(q)
\quad \text{for some } i \in \{1,2,3\} \text{ and some } q \in \T_m.
$$

By Lemma~\ref{three viewpoints}, we can choose a viewpoint $0_i$
which realizes this facet and from which its intersection with $E$
is seen at a good angle.
Formally, among the indices $i_0 \in \{1,2,3\}$ such that
$I = \facet_{P_{i_0} \cap E}(q_0)$ for some $q_0 \in \T_m$,
we choose the one that maximizes the distance from $0$ to the
affine span of the edge
$$
\II = \Facet_{P_{i_0} \cap E}(I).
$$
By Lemma~\ref{three viewpoints},
\begin{equation}                        \label{dist}
\dist(0, \aff(\II)) \ge 1.
\end{equation}
Because only facets of angular length $> 2\pi/m$
were included in the set in \eqref{exp in limit}, we have
$\ang(\II) > 2\pi/m$.
It follows that $\II$ contains some point $q''$ in $\T_m$.

Summarizing, we have realized every facet $I = \facet_{P_i \cap E}(q)$
from \eqref{exp in limit}
as $I = \facet_{P_{i_0} \cap E}(q'')$
for some $i_0$ and some $q'' \in \T_m$.

Recall that when the event $\EE$ holds, all points of $P$ have norm
at most $2$. Thus all points of $P_{i_0}$ have norm at most
$\|0_{i_0}\| + 2 = 10$.
Since $\II \subset P_{i_0}$, all points in $\II$ also have norm at most $10$.
Therefore bound \eqref{dist} yields, by view of Lemma~\ref{angular euclidean},
that {\em the angular and Euclidean distances are equivalent on $\II$ up
to a factor of $c$.}

We shall call a facet (edge) of a polygon nondegenerate
if the angular and Euclidean distances are equivalent on it
up to the factor $c$. We have shown that
$$
\Exp
\le  \lim_{m \to \infty} \E \,
  |\{ \text{nondegenerate } \facet_{P_i \cap E}(q) : \;
  q \in \T_m, \; i=1,2,3 \}| \cdot \one_\EE + 2.
$$
Each facet may correspond to more than one $q$. We are going to leave only one $q$
per facet, namely the $q = q(\theta)$ with the maximal $\theta$ (according
to the parametrization of the torus in the beginning of the argument).
Therefore, the angular distance of such $q$ to one boundary of $\Facet_{P_i \cap E} (q)$
(one of the endpoints of this interval) is at most $2\pi/m$.
The nondegeneracy of this facet then implies that the usual distance
of $q_{P_i}$ to the boundary of $\Facet_{P_i \cap E} (q)$, thus also to the boundary
of $\Facet_{P_i} (q)$, is at most $\frac{1}{c} \cdot 2 \pi/m =: C/m$. Therefore
\begin{align*}
\Exp
&\le  \lim_{m \to \infty} \E \,
  |\{ \facet_{P_i}(q) \ \text{such that}\
    \dist(q_{P_i}, \partial \Facet_{P_i}(q)) \le \frac{C}{m}, \
  q \in \T_m, \; i=1,2,3 \}| \cdot \one_\EE + 2 \\
&\le 3 \max_{i=1,2,3}
\lim_{m \to \infty} \E \,
  |\{ \facet_{P_i}(q) \ \text{such that}\
    \dist(q_{P_i}, \partial \Facet_{P_i}(q)) \le \frac{C}{m}, \
  q \in \T_m\}| \cdot \one_\EE + 2.
\end{align*}

Recall that by \eqref{Pi} and \eqref{0i},
the polytope $P_k$ is the convex hull of the origin and
a translate of the polytope $P$ by a fixed vector of norm $8$.
Therefore
\begin{equation}                                        \label{Exp Exp}
  \Exp \le 3 \max \Exp_0 + 2,
\end{equation}
where
\begin{equation}                                        \label{Exp0}
\Exp_0 = \lim_{m \to \infty} \E \,
  |\{ \facet_P (q) \ \text{such that}\ \dist(q_P, \partial \Facet_P(q)) \le \frac{C}{m}, \
  q \in \T_m\}| \cdot \one_\EE
\end{equation}
and the maximum in \eqref{Exp Exp} is over all centers of
the distributions $a_i$ of norm at most $1+8 = 9$.
Scaling the vectors down by $9$, we can bound $\Exp_0$ using the
following lemma:

\begin{lemma}[Discretized counting]                 \label{DC}
  Let $a_1,\ldots,a_n$ be independent Gaussian vectors in $\R^d$
  with centers of norm at most $1$, and whose standard deviation $\s$ satisfies
  \eqref{ai bi sigma}.
  Then
  $$
  \Exp_0 \le C_0 d^3 \s^{-4},
  $$
  where $C_0$ is an absolute constant.
\end{lemma}

This lemma and \eqref{Exp Exp} complete the proof of Theorem~\ref{section}.

\subsection{Proof of Lemma~\ref{DC}}

As before, $\binom{n}{d}$ will denote
the set of all $d$-subsets of $\{1,\ldots,n\}$. We have
\begin{equation}                        \label{sum sum}
\Exp_0 = \lim_{m \to \infty} \sum_{q \in \T_m} \sum_{I \in \binom{n}{d}}
\P \,
  \{ \facet_P(q) = I
    \text{ and } \dist(q_P, \partial \conv(a_i)_{i \in I}) \le \frac{C}{m}
    \text{ and } \EE \}.
\end{equation}
For every $q$, there exists at most one $I \in \binom{n}{m}$ such that
$\facet_P(q) = I$. Furthermore, the last equation is equivalent to
$q_P \in \conv(a_i)_{i \in I}$. Hence
$$
\sum_{I \in \binom{n}{d}} \P \{ \facet_P(q) = I \}
= \sum_{I \in \binom{n}{d}} \P \{ q_P \in \conv(a_i)_{i \in I} \} \le 1.
$$
Using this bound in \eqref{sum sum}
and estimating $\sum_{q \in \T_m}$ above by
$m \cdot \max_{q \in \T_m}$, we obtain
\begin{equation}                                \label{Exp0 p0}
  \Exp_0 \le \lim_{m \to \infty} m \cdot p_0(m),
\end{equation}
where
$$
p_0(m) = \max_{q \in \T_m, I \in \binom{n}{d}}
  \P \{ \dist(q_P, \partial \conv(a_i)_{i \in I}) \le C/m
    \text{ and } \EE
    \; \big| \;
    q_P \in \conv(a_i)_{i \in I} \}.
$$
Thus we should be looking for an estimate of the type $p_0(m) \lesssim 1/m$.

To this end, we fix an arbitrary point $q \in \T^m$ and a set $I \in \binom{n}{d}$.
Consider the hyperplane
$$
E_I := \Span(a_i)_{i \in I}
$$
and the point
$$
q_I := \text{point where the direction of $q$ pierces the hyperplane $E_I$}.
$$
Then
\begin{equation}                                \label{face is I}
  \facet_P(q) = I \Leftrightarrow
  \begin{cases}
    q_I \in \conv(a_i)_{i \in I}; \\
    \text{all vectors $(a_i)_{i \in I^c}$ are below $E_I$}.
  \end{cases}
\end{equation}

We now pass to the local coordinates $(b_i)_{i \in I}$
for the hyperplane $E_I$, using the change of variables
$$
a_i = R_\w b_i + r \w, \qquad i \in I,
$$
described in Appendix~\ref{a:change}.

We condition on a realization of $r$, $\w$ and the vectors
$(a_i)_{i \in I^c}$. This fixes the hyperplane $E_I$ and the point $q_I$
determined by $r$ and $\w$.
The density of the vectors $(b_i)_{i \in I}$ is given
in Lemma~\ref{change} in Appendix~\ref{a:change}.
By \eqref{face is I}, we can assume that
the vectors $(a_i)_{i \in I^c}$ are below $E_I$.

Let $p \in \R^{d-1}$ be the (fixed) representation of $q_I$ in the new variables,
i.e.
$$
q_I = R_\w p + r \w.
$$
Consider the event
$$
\EE_0 := \{ \text{$\|b_i\| \le 2$, $i=1,\ldots,n$} \}
$$
By part 1 of Lemma~\ref{change},
$$
\EE \subseteq \EE_0.
$$

By \eqref{face is I},
$$
\facet_P(q) = I
\Leftrightarrow q_I \in \conv(a_i)_{i \in I}
\Leftrightarrow p \in \conv(b_i)_{i \in I}.
$$
Moreover, if $p \in \conv(b_i)_{i \in I}$ then
$\|p\| \le \max_{i \in I} \|b_i\|$, thus
$$
\{ \EE \text{ and } \facet_P(q) = I \}
\subseteq \{ \EE_0 \text{ and } p \in \conv(b_i)_{i \in I} \}
\subseteq \{ \|p\| \le 2 \}.
$$
Summarizing, we have shown that
$$
p_0(m) \le
\max \P \{ \dist(p, \partial \conv(b_i)_{i \in I}) \le C/m
    \text{ and } \EE_0
    \; \big| \;
    p \in \conv(b_i)_{i \in I} \},
$$
where the maximum is over all vectors $p \in \R^{d-1}$ such that $\|p\| \le 2$.
We may assume that $p=0$ by translating all the vectors by $-p$.
Before this translation, the densities $\nu_i$ that make up the
density of $(b_i)_{i \in I}$ in \eqref{dens} had centers of norm at most $1$
by Lemma~\ref{change}. After the translation, their norms will be at most
$1 + \|p\| \le 3$. Similarly, the constant $2$ in the definition of
$\EE_0$ will change to $2 + \|p\| \le 4$.

It remains to use the Distance Lemma~4.1.2 of \cite{ST}:

\begin{lemma}[Distance Lemma (Spielman-Teng)]       \label{distance lemma}
  Let $\nu_1,\ldots, \nu_d$ be densities of Gaussian vectors in $\R^{d-1}$
  with centers of norm at most $3$ and with standard deviation
  $\s < 1/3 \sqrt{d \log n}$.
  Then, for every $\e > 0$, we have
  $$
  \P \{ \dist(0, \aff(b_2,\ldots,b_d)) < \e
    \text{ and all } \|b_i\| \le 4 \}
  \le C_1 d^2 \s^{-4} \e,
  $$
  where $(b_1,\ldots,b_d)$ have joint density proportional to
  $$
  |\conv(b_1,\ldots,b_d)| \cdot \prod_{i=1}^d \nu_i(b_i),
  $$
  and $C_1$ is an absolute constant.
\end{lemma}

Note that $\dist(0, \partial \conv(b_i)_{i \in I}) \le C/m$ implies that
there exists $i_0 \in I$ such that
$$
\dist(0, \aff(b_i)_{i \in I - \{i_0\} }) \le C/m.
$$
Since there are $d$ choices for $i_0$, the Distance Lemma~\ref{distance lemma}
yields:
$$
p_0(m) \le d \cdot C_1 d^2 \s^{-4} (C/m) \le C_2 d^3 \s^{-4} / m,
$$
where $C_2 = C_1 C$.
By \eqref{Exp0 p0}, we conclude that
$$
\Exp_0 = O(d^3 \s^{-4}).
$$
This completes the proof. \endproof

\section{Sections of random polytopes with an added facet} \label{s:section+}

In this section, we prove Corollary~\ref{section+}
about the size of planar sections of random polytopes with an added facet.

The main difficulty is that not all vectors $(a_1,\ldots,a_{n+d})$ are independent.
The last $d$ vectors correlate with the first $n$ vectors through the random variable
$M_0 = e^{\lceil \log M \rceil}$, where
$M = \max_{i=1,\ldots,n} \|a_i\|$.

This difficulty will be resolved similarly to \cite{ST}.
We will show that with high probability,
$M_0$ takes values in a set of cardinality $O(\log \log n)$.
For each fixed $M_0$ in this set, all the vectors $a_1,\ldots,a_{n+d}$
are independent, so we will be able to use Theorem~\ref{section} to get the
desired size of the section $P \cap E$.

\subsection{Boundedness of the random polytope}

Consider
$$
\bar{M} = \max_{i=1,\ldots,n} \|\bar{a}_i\|, \qquad
M = \max_{i=1,\ldots,n} \|a_i\|.
$$
Recall that by \eqref{ai bi sigma},
$\bar{M} \le 1$.

\begin{lemma}                           \label{M bar M}
  Consider the event
  \begin{equation}                      \label{M bdd}
    \EE_1 := \Big\{
      \frac{c_1}{\sqrt{\log n}} \big( \bar{M} + \s \sqrt{d \log n} \big)
      \le M
      \le \bar{M} + 3\s \sqrt{d \log n}
    \Big\},
  \end{equation}
  where $c_1= c/9$, and where $c$ is the absolute constant in Lemma~\ref{gaussian}.
  Then
  $$
  \P \{ \EE_1^c \} \le \binom{n}{d}^{-1},
  $$
\end{lemma}

\begin{proof}
{\em Upper bound.} \
By part 1 of Lemma~\ref{gaussian},
$$
\P \{ |M - \bar{M}| > 3\s \sqrt{d \log n} \}
  \le \{ \max_{i=1,\ldots,n} \|a_i - \bar{a}_i\| > 3 \s \sqrt{d \log n} \}
  \le n^{-2.9d+1} \le 0.0015 \binom{n}{d}^{-1}.
$$

{\em Lower bound.} \
We consider two cases. If $\bar{M} \ge 8 \s \sqrt{d \log n}$ then,
using the last estimate, we have:
$$
\P \{ M < \frac{1}{2} ( \bar{M} + \s \sqrt{d \log n} ) \}
\le \P \{ M < \bar{M} - 3 \s \sqrt{d \log n} \}
\le 0.0015 \binom{n}{d}^{-1}.
$$
If $\bar{M} < 8 \s \sqrt{d \log n}$ then, using part 2 of Lemma~\ref{gaussian}
and the independence of $a_i$, we obtain
$$
\P \big\{ M < \frac{c_1}{\sqrt{\log n}} ( \bar{M} + \s \sqrt{d \log n} ) \big\}
\le \P \{ M < 9 c_1 \s \sqrt{d} \}
\le (e^{-d})^n
\le 0.5 \binom{n}{d}^{-1}.
$$
This completes the proof.
\end{proof}

\subsection{Discretization of $M$}

Recall that $M_0 = e^{\lceil \log M \rceil}$.
It follows from Lemma~\ref{M bar M} that $M_0$ is likely to take
values in a small set.

Indeed, if $\EE_1$ holds, then $M_0$ takes values in a (non-random) set
$$
\MM_0 := \{ e^{\lceil \log M \rceil} : \;
   \text{$M$ satisfies the inequalities in \eqref{M bdd}} \}
$$
of cardinality $|\MM_0| = O(\log \log n)$.

Therefore, using as before that the number of edges of the polytope
$P \cap E$ is bounded by $\binom{n}{d}$, we have:
\begin{align}           \label{E0}
\E |\edges(P \cap E)|
  &\le \E |\edges(P \cap E)| \cdot \one_{\EE_1} + 1 \notag\\
  &= \sum_{m \in \MM_0} \E |\edges(P \cap E)| \cdot \one_{\{M_0 = m\}} + 1 \notag\\
  &\le C \log \log n \cdot \max_{M_0 \in \MM_0} \E_0 |\edges(P \cap E)| + 1,
\end{align}
where $\E_0$ denotes the expectation with respect to the distribution
of $(a_1,\ldots,a_{n+d})$ as in the statement of Corollary~\ref{section+}
{\em except that the value of $M_0 \in \MM_0$ is fixed}.
According to this new distribution, all vectors $a_1,\ldots,a_{n+d}$
are now independent.

\subsection{Vertices with different standard deviations}

Now we dilate the vectors $a_i$ so that the centers of their distributions
become at most $1$. To this end, define
$$
b_i := \frac{a_i}{C' M_0 \sqrt{\log n}}, \qquad i = 1,\ldots, n+d,
$$
where $C' = 3/c_1$, where $c_1$ is the constant in Lemma~\ref{M bar M}.
Let us estimate the standard deviations and the centers of the distributions
of $b_i$.

{\em Vectors $b_1, \ldots, b_n$.} \
The standard deviation of $a_1,\ldots,a_n$ is $\s$.
Hence the standard deviation of $b_1,\ldots,b_n$ is
$$
\s' := \frac{\s}{C' M_0 \sqrt{\log n}}.
$$
Using the lower bound in \eqref{M bdd} (which holds by the
definition of $\MM_0$) and the trivial inequality $M \le M_0$,
we conclude that the standard deviation satisfies a condition
of the type \eqref{ai bi sigma}, namely
\begin{equation}                    \label{sigma' above}
  \s' \le \frac{\s}{C' M \sqrt{\log n}}
  \le \frac{\s}{C' c_1 \s \sqrt{d \log n}}
  \le \frac{1}{3 \sqrt{d \log n}}.
\end{equation}
On the other hand, since $\bar{M} \le 1$ and by the upper bound in \eqref{M bdd},
we have $M_0 \le 3M \le 3 \bar{M} + 9 \s \sqrt{d \log n} \le 3 + 9/6 \le 5$.
Thus
\begin{equation}                    \label{sigma' below}
  \s' \ge \frac{\s}{5 C' \sqrt{\log n}}.
\end{equation}
The centers of the distributions of $a_1,\ldots,a_n$ have norms at most $\bar{M}$.
Since $M \le M_0$ and using the lower bound in \eqref{M bdd},
we conclude that the centers of the distributions of $b_1,\ldots,b_n$ have
norms at most
$$
\frac{\bar{M}}{C' M_0 \sqrt{\log n}}
\le \frac{M}{c_1 C' M_0} \le 1.
$$

{\em Vectors $b_{n+1}, \ldots, b_{n+d}$.} \
The standard deviation of $a_{n+1},\ldots,a_{n+d}$ is $2 M_0 \s_1$.
Hence the standard deviation of $b_{n+1},\ldots,b_{n+d}$ is
\begin{equation}                    \label{sigma''}
  \s'' := \frac{2\s_1}{C' \sqrt{\log n}}.
\end{equation}
Since $\s_1$ satisfies \eqref{ellsigma} and $C' \ge 1$,
the standard deviation $\s''$ satisfies a condition
of the type \eqref{ai bi sigma}, namely
\begin{equation}                    \label{sigma'' above}
  \s'' \le \frac{1}{3 \sqrt{d \log n}}.
\end{equation}
The centers of the distributions of $a_{n+1},\ldots,a_{n+d}$
have norms at most $3 M_0$.
Hence the centers of the distributions of $b_{n+1},\ldots,b_{n+d}$ have
norms at most
$$
\frac{3}{C' \sqrt{\log n}} \le 1.
$$

Now we apply Lemma 4.3.2 from \cite{ST} which allows us to
reduce standard deviations of the vertices to one (minimum) value.

\begin{lemma}[Gaussians free]                        \label{free}
  Let $d_1,\ldots,d_n$ be independent Gaussian vectors in $\R^d$
  with standard deviations $\s_1,\ldots,\s_n$. Let $\s_0 > 0$.
  Assume that $\s_0 \le \s_i \le 1/3 \sqrt{d \log n}$ for all $i$.
  Let $E$ be a plane in $\R^d$. Then the random polytope
  $P = \Conv(d_1,\ldots,d_n)$ satisfies
  $$
  \E \, |\edges (P \cap E)| \le D(d,\s_0) + 1.
  $$
  where $D(d,\s)$ denotes the right hand side of \eqref{section bound}.
\end{lemma}

We use this lemma for the vectors $b_1,\ldots,b_{n+d}$
and for
$$
\s_0 := \min \Big( \frac{\s}{5 C' \sqrt{\log n}},
\frac{2\s_1}{C' \sqrt{\log n}} \Big)
\ge c \log^{-1/2}(n) \cdot \min(\s,\s_1),
$$
where $c > 0$ is some absolute constant.

Indeed, it follows from \eqref{sigma' below} and \eqref{sigma''}
that $\s_0 \le \min(\s',\s'')$.
Similarly, \eqref{sigma' above} and \eqref{sigma'' above}
state that $\max(\s',\s'') \le 1/3 \sqrt{d \log n}$.
Thus Lemma~\ref{free} applies, and it yields
$$
\E_0 |\edges(P \cap E)| \le D(d, \s_0) + 1.
$$
Using this in \eqref{E0}, we conclude that
$$
\E |\edges(P \cap E)| \le C \log \log n \cdot D(d, \s_0) + 1.
$$
This completes the proof of Corollary~\ref{section+}.

\bigskip

\renewcommand{\thesection}{\Alph{section}}
\setcounter{section}{0}

\section{Appendix \thesection. Proof of Proposition~\ref{interpolation}}         \label{a:interpolation}

In this section, we prove Proposition~\ref{interpolation},
which allows us to interpolate between an arbitrary linear program
and a unit linear program.

For a fixed $t \in [0,1]$, let (LP$_t$) denote the interpolation
program (Int LP) with this fixed value of $t$.
The feasible sets (polytopes) of (LP) and of (LP$_t$) will be
denoted by $P$ and $P_t$ respectively. They are subsets of $\R^d$.

\subsection{Recession cone}

Our proof of (i) of Proposition~\ref{interpolation}
will be based on an analysis of the {\em recession cone},
defined as
$$
\Recess(LP) = \Recess(P)
:= \{ x : \; Mx+P \subseteq P \text{ for all } M \ge 0 \}.
$$
The polytope $P$ is unbounded iff its recession cone is nonempty.

\begin{lemma}[Recession cone]                   \label{recess}
  Assume (LP) is feasible. Then
  $$
  \Recess(\LP) = \{x :\; Ax \le 0\} = \big( \cone(a_i)_{i=1}^n \big)^\circ
  $$
\end{lemma}

\begin{proof}
The second equation is trivial.

To prove the inclusion $\Recess(\LP) \subseteq \{x :\; Ax \le 0\}$,
let us fix arbitrary $x \in \Recess(P)$ and $x_0 \in P$.
By the definition of the recession cone,
$Mx + x_0 \in P$ for all $M>0$.
Since $P$ is the feasible polytope of (LP), we have
$$
A(Mx + x_0) \le b \qquad \text{for all $M>0$}.
$$
Hence
$$
Ax \le \frac{1}{M} (b - A x_0)
\to 0 \qquad \text{as $M \to \infty$}.
$$
It follows that $Ax \le 0$.

To prove the inclusion $\Recess(\LP) \supseteq \{x :\; Ax \le 0\}$,
let us fix $x$ such that $Ax \le 0$ and $x_0 \in P$.
Since $P$ is the feasible polytope of (LP), we have $A x_0 \le b$.
Then for every $M \ge 0$ we have
$$
A(Mx + x_0) = M \cdot Ax + Ax_0 \le b.
$$
Thus $Mx + x_0 \in P$ for every $M \ge 0$. Hence $x \in \Recess(P)$.
This completes the proof.
\end{proof}

\begin{lemma}[Boundedness of LP]                    \label{boundedness}
  Assume the linear program (LP) is feasible. Then
  (LP) is bounded iff
  $$
  z \in \big( \Recess(LP) \big)^\circ = \cone(a_i)_{i=1}^n.
  $$
\end{lemma}

\begin{proof}
{\em Necessity.}
Assume $z \not\in \big( \Recess(LP) \big)^\circ$. Then there exists
$x \in \Recess(LP)$ such that $\< z,x \> > 1$.
Let us fix an arbitrary $x_0 \in P$.
By the definition of the recession cone,
$Mx + x_0 \in P$ for every $M \ge 0$.
Therefore, the vectors $Mx + x_0$ are feasible for (LP).
On the other hand, the objective function can take arbitrarily
large values on such vectors:
$$
\< z, Mx + x_0 \> = M \< z,x \> + \< z,x_0 \> \to \infty
\qquad \text{as $M \to \infty$}.
$$
Thus (LP) is unbounded.

{\em Sufficiency.}
Assume $z \in \cone(a_i)_{i=1}^n$.
Then we can write
$z = \sum_{i=1}^n \l_i a_i$ for some $\l_i \ge 0$.
Let $x$ be any feasible vector for (LP), that is $Ax \le b$.
Then, the objective function is bounded as
$$
\< z,x \> \le \sum_{i=1}^n \l_i \< a_i, x\> \le \sum_{i=1}^n \l_i b_i.
$$
Thus (LP) is bounded.
This completes the proof.
\end{proof}

\begin{remark}
  The main point in both lemmas is that their conclusion does
  not depend on the right hand side $b$ of (LP).
  Thus, {\em for feasible linear programs, their boundedness does
  not depend on the right hand side $b$.}
\end{remark}

\subsection{Proof of Proposition~\ref{interpolation}}

\paragraph{\bf Proof of (i)} \
Since unbounded linear programs are feasible, Lemma~\ref{boundedness}
implies that (LP) is unbounded iff (\UnitLP) is unbounded.

Next, if (\UnitLP) is unbounded then (\IntLP) is clearly unbounded
for every $\l$. Indeed, the feasible set of (\IntLP) contains the
$P_0$ as a section (for $t=0$), and $P_0$ is the feasible set of (\UnitLP).

The remaining part of (i) is: if (\IntLP) is unbounded for some $\l$,
then (\UnitLP) is unbounded. We shall now prove this.

The unboundedness of (\IntLP) means that for every $M \ge 1$ there
exists a feasible point $(x_M, t_M)$ for (\IntLP) such that
\begin{equation}                            \label{xM tM}
  \< z, x_M \> + \l t_M \to \infty
  \qquad \text{as $M \to \infty$}.
\end{equation}
Since $t_M \in [0,1]$, the definition of (\IntLP) yields
\begin{equation}                            \label{RHS bdd}
  Ax_M \le t_M b + (1-t_M) \one
    \le \max(\|b\|_\infty, 1) \cdot \one
    =: B \cdot \one.
\end{equation}
Writing this estimate as $A \big( \frac{1}{B} x_M \big) \le \one$,
we see that the vectors $\frac{1}{B} x_M$ are feasible for (\UnitLP)
for all $M \ge 1$.
On the other hand, the objective function can take arbitrarily large
values on such vectors. Indeed, since $t_M \in [0,1]$, estimate
\eqref{xM tM} implies that
$$
\< z, \frac{1}{B} x_M \> \to \infty
\qquad \text{as $M \to \infty$}.
$$
Thus (\UnitLP) is unbounded.

\paragraph{\bf Proof of (iii)} \
Assume that (LP) is not unbounded.

{\em Sufficiency.}
Assume (LP) is feasible. Then the interpolation (LP$_t$) is
clearly feasible for every $0 \le t \le 1$.
Indeed, if $x$ is feasible for (LP), then $tx$ is feasible for (LP$_t$).
Similarly to \eqref{RHS bdd}, we have that $Ax \le B \cdot \one$
fr all feasible points $x$ of (LP$_t$). In terms of the feasible
polytopes, this means that $P_t \subseteq B \cdot P_0$.

Since (LP) is not unbounded by the assumption, part (i) implies
that (\UnitLP) is not unbounded. Being always feasible ($0$ is a
feasible set), (\UnitLP) must therefore be bounded.
Hence
\begin{equation}                        \label{lambda0}
  \big| \max_{x \in P_t} \< z,x \> \big|
  \le B \cdot \big| \max_{x \in P_0} \< z,x \> \big|
  =: \l_0.
\end{equation}

We can write (\IntLP) as the optimization problem
$$
\max_{t \in [0,1]} f(t),
\quad \text{where }
f(t) = \max_{x \in P_t} \frac{1}{\l} \< z,x \> + t.
$$
By \eqref{lambda0}, we have
$$
|f(t) - t| \le \l_0/\l
\quad \text{for all $t \in [0,1]$}.
$$
It then follows that
$$
\argmax_{t \in [0,1]} f(t) \ge 1 - 2\l_0/\l,
$$
which converges to $1$ as $\l \to \infty$.

We have shown that a solution $(x_\l, t_\l)$ of (\IntLP)
with parameter $\l$ satisfies $t_l \to 1$ as $\l \to \infty$.
On the other hand, $(x_\l, t_\l)$ is a vertex of the feasible
polytope of (\IntLP). This polytope does not depend on $\l$,
thus there are finitely many choices for $(x_\l, t_\l)$.
It follows that $t_\l = 1$ for all sufficiently large $\l$.
The sufficiency in (iii) is proved.

{\em Necessity.} This part is trivial. Indeed, if $(x,1)$ is a solution
of (\IntLP) with some parameter $\l$,
then $x$ is a solution of (LP$_1$) and thus $x$
is a solution of (LP).

\paragraph{\bf Proof of (ii)} \
Assume (LP) is not unbounded.

Note that for all sufficiently small $t$, namely
for $t \in [0,t_0]$ with $t_0 := 1/\|b-\one\|_\infty$,
the right hand side of (\IntLP) is nonnegative:
$$
tb + (1-t)\one \ge 0.
$$
Hence (LP$_t$) is feasible for all $t \in [0,t_0]$
($0$ is a feasible point).

For a fixed $\l < 0$, we can write (\IntLP) as the optimization problem
$$
\max_{t \in \Dom(g)} g(t),
\quad \text{where }
g(t) = \max_{x \in P_t} \frac{1}{-\l} \< z,x \> - t.
$$
The domain $\Dom(g)$ consists of all $t \in [0,1]$ for which
$P_t$ is nonempty, i.e. (LP$_t$) is feasible.
In particular, $[0,t_0] \subseteq \Dom(g)$.

By \eqref{lambda0}, we have
$$
|g(t) + t| \le \l_0/(-\l)
\quad \text{for all $t \in \Dom(g)$}.
$$
>From this and from the fact that $\Dom(g)$ contains a
neighborhood $[0,t_0]$ of $0$ it follows that
$$
\argmax_{t \in \Dom(g)} g(t) \le - 2\l_0/\l,
$$
which converges to $0$ as $\l \to -\infty$.

We have shown that a solution $(x_\l, t_\l)$ of (\IntLP)
with parameter $\l$ satisfies $t_l \to 0$ as $\l \to -\infty$.
Similarly to our proof of (iii), we deduce that
$t_\l = 0$ for all sufficiently small $\l$.
This proves (ii).

\paragraph{\bf Proof of (iv)} \
This follows directly from (iii) since (\IntLP) for $t=1$
is (LP).

The proof of Proposition~\ref{interpolation} is complete.

\section{Appendix \thesection. Proof of Theorem~\ref{adding constraints}}     \label{a:adding constraints}

\subsection{Part~1}
We need to prove that (4a) and (4b) in \textsc{Adding Constraints}
imply that in the polytope $P^+ = \Conv(0,a_1,\ldots,a_{n+d})$,
one has $\facet(z) = \{n+1,\ldots,n+d\}$.
By (4a), it will be enough to show that all points $a_1,\ldots,a_n$
lie below the affine span $\aff(a_{n+1},\ldots,a_{n+d}) =: H$.
Since all these points have norm at most $M$, it will suffice
to show that all vectors $x$ of norm at most $M$ are below $H$.
By (4b), the normal $h$ to $H$ has norm at most $1/M$, thus
$\< h, x\> \le 1$. Thus $x$ is indeed below $H$.
This completes the proof.

\subsection{Part~2}
Without loss of generality, we can assume that $M_0 = M$.
Also, by the homogeneity, we can assume that $M = 1/2$.
Thus there is no dilation in step~2 of \textsc{Adding Constraints}.
Let $H$ be a numb half-space.
It suffices to show that
\begin{gather}
  \P \{ z_0 \in \cone(a_{n+1},\ldots,a_{n+d}) \} \ge 0.99;  \label{in cone} \\
  \P \{ \text{normal $h$ to $\aff(a_{n+1},\ldots,a_{n+d})$
              satisfies $\|h\| \le \frac{1}{M}$} \} \ge 0.99;   \label{normal small} \\
  \P \{ \text{$a_{n+1},\ldots,a_{n+d}$ are in $H$} \} \ge 1/3.  \label{in numb}
\end{gather}

Events in \eqref{in cone} and \eqref{normal small} are invariant under
the rotation $U$. So, in proving these two estimates we can assume that
$U$ is the identity, which means that $z_0 = z'_0$ and
$\bar{a}_i = \bar{a}'_i$ for $i = n+1,\ldots,n+d$.
We can also assume that $d$ is bigger than some suitable absolute
constant ($100$ will be enough).

We will use throughout the proof the known estimate on the spectral norm
of random matrices.
Recall that the spectral norm of a $d \times d$ matrix $B$ is defined as
$$
\|B\| = \max_{x \in \R^d} \frac{\|Bx\|}{\|x\|}.
$$
For a random $d \times d$ matrix $B$ with independent standard normal random entries,
one has
\begin{equation}                                            \label{random matrix norm}
  \P \{ \|B\| > 2 t \sqrt{d} \} \le 2^d (d-1) t^{d-2} e^{-d (t^2-1)/2}
  \ \ \ \text{for $t \ge 1$}
\end{equation}
see e.g. \cite{Sz}. Therefore, for a random $d \times d$ matrix $G$
with independent entries of mean $0$ and variance $\s_1$, we have
$$
\P \{ \|G\| > 2 \s_1 t \sqrt{d} \} \le 2^d (d-1) t^{d-2} e^{-d (t^2-1)/2}
\ \ \ \text{for $t \ge 1$}.
$$
(this follows from \eqref{random matrix norm} for $B=G/\s_1$).

In particular,
\begin{equation}                        \label{norm G}
\P \{ \|G\| \le 3 \s_1 \sqrt{d} \} \ge 0.99.
\end{equation}

We will view the vectors $a_{n+1},\ldots,a_{n+d}$ as images of some fixed orthonormal
vector basis $e_{n+1},\ldots,e_{n+d}$ of $\R^d$. Denote
$$
\one = \sum_{i=n+1}^{n+d} e_i.
$$
We define the linear operator $T$ in $\R^d$ so that
$$
\bar{a}_i = T e_i, \ \ \ a_i = (T+G) e_i, \ \ \ i = n+1,\ldots,n+d.
$$
We first show that
\begin{equation}                        \label{norm T}
  \|T^{-1}\| \le 1/\ell.
\end{equation}
Indeed, $\conv(e_{n+1},\ldots,e_{n+d})$ is a simplex with
center $d^{-1} \one$ of norm $d^{-1/2}$ and radius
$\|d^{-1} \one - e_i\| = \sqrt{1-1/d}$.
Similarly, $\conv(\bar{a}_{n+1},\ldots,\bar{a}_{n+d})$ is a simplex with
center $z_0$ of norm $1$ and radius $\|z_0 - \bar{a}_i\| = \ell$.
Therefore we can write $T = V T_1$ with a suitable $V \in O(d)$, and where
$T$ acts as follows: if $x = x_1 + x_2$ with $x_1 \in \Span(\one)$
and $x_2 \in \Span(\one)^\perp$, then
$T_1 x = d^{1/2} x_1 + \ell (1 - 1/d)^{-1/2} x_2$.
Thus $\|T^{-1}\| = \|T_1^{-1}\| = \ell^{-1} (1 - 1/d)^{1/2} \le 1/\ell$.
This proves \eqref{norm T}.

\subsubsection{Proof of \eqref{in cone}}
An equivalent way to state \eqref{in cone} is that
\begin{equation}                        \label{z0 sum}
  z_0 = \sum_{i=n+1}^{n+d} c_i a_i \ \ \
  \text{where all $c_i \ge 0$}.
\end{equation}
Recall that $a_i = (T+G) e_i$. Multiplying both sides of \eqref{z0 sum}
by the operator $(T+G)^{-1}$, we obtain
$$
(T+G)^{-1} z_0 = \sum_{i=n+1}^{n+d} c_i e_i.
$$
Taking the inner product of both sides of this equation with the vectors $e_i$,
we can compute the coefficients $c_i$ as
\begin{equation}                        \label{coeff}
c_i = \< (T+G)^{-1} z_0, e_i \> .
\end{equation}
On the other hand, $z_0$ is the center of the simplex
$\conv(\bar{a}_{n+1},\ldots,\bar{a}_{n+d})$, so
$z_0 = \sum_{i=n+1}^{n+d} (1/d) \bar{a}_i$.
Since $\bar{a}_i = T e_i$, a similar argument shows that
\begin{equation}                        \label{coeff'}
\frac{1}{d} = \< T^{-1} z_0, e_i \> .
\end{equation}
Thus to bound $c_i$ below, it suffices to show that the right
sides of \eqref{coeff} and \eqref{coeff'} are close.
To this end, we use the identity
$T^{-1} - (T+G)^{-1} = (1 + T^{-1}G)^{-1} T^{-1} G T^{-1}$
and the estimate $\|1 + S\| \le (1-\|S\|)^{-1}$ valid for operators
of norm $\|S\|<1$. Thus the inequality
\begin{equation}                        \label{operator stability}
\|(T+G)^{-1} - T^{-1}\| \le \frac{\|T^{-1}\|^2 \; \|G\|}{1 - \|T^{-1}\| \|G\|}
\le \frac{1}{2d}
\end{equation}
holds with probability at least $0.99$,
where the last inequality follows from \eqref{norm G}, \eqref{norm T} and
from our choice of $\ell$ and $\s_1$ made in \eqref{ellsigma}.
Since $z_0$ and $e_i$ are unit vectors, \eqref{operator stability} implies
that the right sides of \eqref{coeff} and \eqref{coeff'} are within $\frac{1}{2d}$
from each other. Thus $c_i \ge \frac{1}{2d} > 0$ for all $i$.
This completes the proof of \eqref{in cone}.

\subsubsection{Proof of \eqref{normal small}}
We claim that the normal $z'_0$ to $\aff(\bar{a}_{n+1},\ldots,\bar{a}_{n+d})$ and
the normal $h$ to $\aff(a_{n+1},\ldots,a_{n+d})$ can be computed as
\begin{equation}                    \label{normals}
  z'_0 = (T^*)^{-1} \one, \ \ \
  h = ((T+G)^*)^{-1} \one.
\end{equation}
Indeed, for every $i \in \{n+1,\ldots,n+d\}$ we have
$$
\< (T^*)^{-1} \one, \bar{a}_i \>
= \< \one, T^{-1} \bar{a}_i \>
= \< \one, e_i \>
= 1.
$$
Hence by the definition of the normal, the vector $(T^*)^{-1} \one$ is
the normal to $\aff(\bar{a}_{n+1},\ldots,\bar{a}_{n+d})$.
The second identity in \eqref{normals} is proved in a similar way.

Since $z'_0$ is a unit vector, to bound the norm of $h$ it suffices
to estimate
$$
\|h - z'_0\| \le \| ((T+G)^*)^{-1} -  (T^*)^{-1}\| \|\one\|
 = \| (T+G)^{-1} -  T^{-1}\| \|\one\|.
$$
By \eqref{operator stability} and using $\|\one\| = d^{-1/2}$,
with probability at least $0.99$ one has
$\|h - z'_0\| \le \frac{1}{2} d^{-3/2} \le 1$.
Thus $\|h\| \le 2$, which completes the proof of \eqref{normal small}.

\subsubsection{Proof of \eqref{in numb}}
Let $\nu$ be a unit vector such that the half-space is
$H = \{ x :\; \< \nu, x\> \ge 0\}$.
Then \eqref{in numb} is equivalent to saying that
$$
\P \{ \< \nu, a_i \> \ge 0, \ \ i=n+1,\ldots,n+d \} \ge 1/3.
$$
We will write
\begin{equation}                    \label{three terms}
\< \nu, a_i \>
=  \< \nu, z_0 \> + \< \nu, \bar{a}_i - z_0 \> + \< \nu, a_i - \bar{a}_i \>
\end{equation}
and estimate each of the three terms separately.

Since $z_0$ is a random vector uniformly distributed on the sphere $S^{d-1}$,
a known calculation of the measure of a spherical cap (see e.g. \cite{L} p.25)
implies that
\begin{equation}                    \label{first term}
\P \Big\{ \< \nu, z_0 \> \ge \frac{1}{60 \sqrt{d}} \Big\}
\ge \frac{1}{2} - 0.1.
\end{equation}
This takes care of the first term in \eqref{three terms}.

To bound the second term, we claim that
\begin{equation}                    \label{second term}
\P \Big\{ \max_{i=n+1,\ldots,n+d} |\< \nu, \bar{a}_i - z_0 \> |
   \le \frac{1}{120 \sqrt{d}} \Big\}
\ge 0.99.
\end{equation}
To prove this, we shall use the
rotation invariance of the random rotation $U$. Without changing its
distribution, we can compose $U$ with a further rotation in the hyperplane
orthogonal to $Uz'_0$. More precisely, $U$ is distributed identically
with $VW$. Here $W \in O(d)$ is a random rotation; denote
$z_0 := W z'_0$. Then $V$ is a random rotation in $L = \Span(z_0)^\perp$
and for which $L$ is an invariant subspace, that is $Vz_0 = z_0$.

Then we can write $\bar{a}_i - z_0 = V \ell_i$,
where $\ell_i := W (\bar{a}'_i - z'_0) = W \bar{a}'_i - z_0$.
The vectors $\ell_i$ are in $L$ because
$\< \ell_i, z_0 \> = \< W (\bar{a}'_i - z'_0), W z'_0 \>
= \< \bar{a}'_i - z'_0, z'_0 \> = 0$
since $z'_0$ is a unit vector and, moreover, the normal of $\aff(\bar{a}_i)$.
Since $L$ is an invariant subspace of $V$, it follows that $V \ell_i \in L$.
Furthermore, $\|\ell_i\| = \|\bar{a}'_i - z'_0\| = \ell$.

Let $P_L$ denote the orthogonal projection onto $L$. Then $P_L \nu$
is a vector of norm at most one, so denoting $\nu' = P_L \nu / \|P_L \nu\|$
we have
$$
|\< \nu, \bar{a}_i - z_0 \> |
= |\< \nu, V \ell_i \> |
= |\< P_L \nu, V \ell_i \> |
= |\< V^* P_L \nu, \ell_i \> |
\le |\< V^* \nu', \ell_i \> |
$$
$V^* \nu'$ is a random vector uniformly distributed on the sphere of $L$,
and $\ell_i$ are fixed vectors in $L$ of norm $\ell$.

Then to prove \eqref{second term} it suffices to show that
for $x$ uniformly distributed on $S^{d-2}$ and for any fixed
vectors $\ell_1,\ldots,\ell_d$ in $\R^{d-1}$ of norm $\ell$, one has
\begin{equation}                    \label{mean width}
\P \Big\{ \max_{i=1,\ldots,d} |\< x, \ell_i \> | \le \frac{1}{120 \sqrt{d}} \Big\}
\ge 0.99.
\end{equation}
This is well known as the estimate of the mean width of the simplex.
Indeed, for any choice of unit vectors $h_1,\ldots,h_d$ in $\R^{d-1}$
and any $s > 0$,
$$
\P \Big\{ \max_{i=1,\ldots,d} |\< x, h_i \> | > \frac{s}{\sqrt{d}} \Big\}
\le \sum_{i=1}^d \P \Big\{ |\< x, h_i \> | > \frac{s}{\sqrt{d}} \Big\}
$$
and each probability in the right hand side is bounded by
$p := \exp(-(d-3)^2 s^2 / 4d)$ by the concentration of measure on the sphere
(see \cite{L} (1.1)). We apply this for $h_i = \frac{1}{\ell} \ell_i$
and with $s = \frac{1}{120 \ell}$, which makes $p \le \frac{1}{100d}$.
This implies \eqref{mean width} and, ultimately, \eqref{second term}.

To estimate the third term in \eqref{three terms}, we can condition
on any choice of $U$, so that $\bar{a}_i$ become fixed.
Then $g_i = - \< \nu, a_i - \bar{a}_i \> $ are independent Gaussian
random variables with mean $0$ and variance $\s_1 \le \frac{1}{120 \sqrt{d}} =: s$.
Then
$$
\P \{g_1 > s \} \le \frac{1}{\sqrt{2 \pi}} \exp( -s^2 / 2\s_1^2)
\le \frac{1}{100d}
$$
by a standard estimate on the Gaussian tail and by our choice of $\s_1$ and $s$.
Hence
\begin{equation}                        \label{third term}
\P \Big\{ \min_{i=n+1,\ldots,n+d} \< \nu, a_i - \bar{a}_i \>
  \ge -\frac{1}{120 \sqrt{d}}  \Big\}
= 1 - \sum_{i=n+1}^{n+d} \P \{ g_i > s \}
\ge 0.99.
\end{equation}

Combining \eqref{first term}, \eqref{second term} and \eqref{third term},
we can now estimate \eqref{three terms}:
$$
\P \{ \< \nu, a_i \> \ge 0,  \ \ i=n+1,\ldots,n+d\}
\ge \frac{1}{2} - 0.1 - 0.01 - 0.01 > \frac{1}{3}.
$$
This completes the proof of Theorem~\ref{adding constraints}.
\endproof

\section{Appendix \thesection. Change of variables}     \label{a:change}

The following basic change of variables is useful when
dealing with the hyperplane $E$ spanned by linearly independent
$a_1,\ldots, a_d$ in $\R^d$. It is explained in more detail
in \cite{ST}.

We specify $E$ by choosing $r \in \R_+$ and $\w \in S^{d-1}$
in such a way that $\< \w, a_i \> = r$ for all $i$.
Thus $\w$ is the unit vector in the direction orthogonal to $E$,
and $r$ is the distance from the origin to $E$.

We choose a reference unit vector $h$ in $\R^d$.
The hyperplane $H$ orthogonal to $h$ will be identified with $\R^{d-1}$.
For every $\w \ne -h$, we denote by $R_\w$ the linear transformation
that rotates $h$ to $\w$ in the two-dimensional subspace through $h$ and $\w$
and that is the identity in the orthogonal subspace.
Then one can map a point $b \in \R^{d-1}$ to $a \in E$ by
$a = R_\w b + r\w$.

Let $b_1, \ldots, b_n$ be vectors in $\R^{d-1}$,
and $a_1, \ldots, a_n$ be the corresponding vectors in $\R^d$ under this
change of variables, i.e.
\begin{equation}                    \label{b to a}
  a_i = R_\w b_i + r\w, \qquad i=1,\ldots,d.
\end{equation}
The Jacobian of this change of variables
$(\w,r,b_1,\ldots,b_d) \mapsto (a_1,\ldots,a_d)$ equals
$$
(d-1)! \Vol(\conv(b_1,\ldots,b_d)).
$$
This is a well known formula in the integral geometry due to Blaschke
(see \cite{San}).

\begin{lemma}[Change of variables]                      \label{change}
  Let $a_1,\ldots,a_d$ be independent Gaussian vectors in $\R^d$
  with centers of norm at most $1$ and standard deviation $\s$.
  Let $E = \aff(a_1,\ldots,a_d)$.
  Then the vectors $b_1,\ldots,b_d$ obtained by change of variables
  \eqref{b to a} satisfy the following.

  1. $\|b_i\| \le \|a_i\|$ for all $i$.

  2. Let us condition on a realization of $r$ and $\w$ (i.e. on $E$).
  Then the density of $(b_1, \ldots, b_d)$ is proportional to
  \begin{equation}                  \label{dens}
    \Vol(\conv(b_1,\ldots,b_d)) \, \prod_{i=1}^d \nu_i(b_i),
  \end{equation}
  where $\nu_i$ are the densities of Gaussian vectors in $\R^{d-1}$
  with centers of norm at most $1$ and standard deviation $\s$.
\end{lemma}

\begin{proof}
1. First note that $a_i-r\w$ is orthogonal to $\w$. Indeed,
$\< a_i - r\w, \w \> = \< a_i,\w \> - r = 0$. Therefore
$$
\|a_i\|^2 = \|a_i - r\w\|^2 + \|r\w\|^2
= \|R_\w b_i\|^2 + r^2 = \|b_i\|^2 + r^2.
$$
This proves part 1.

2. The density of $(a_1,\ldots,a_d)$ is $\prod_{i=1}^d \mu_i(a_i)$,
where $\mu_i$ are Gaussian densities. Denote the center of the
distribution of $a_i$ by $\bar{a}_i$. Let $P_E$ denote the orthogonal
projection in $\R^d$ onto $E$.

Since the realization of $E$ is fixed,
the induced distribution $\mu_i (a_i | \, a_i \in E)$
is a distribution of a Gaussian vector in $E$ with center $P_E \bar{a}_i$
and standard deviation $\s$. (See e.g. Proposition~2.4.3 in \cite{ST}.)

Since the change of variables \eqref{b to a} is an isometry $\R^{d-1} \to E$,
part 2 is proved except the upper bound $1$ the norms of the centers of $b_i$.
These centers, denoted by $\bar{b}_i$, are the vectors in $\R^{d-1}$ that
correspond to $P_e \bar{a}_i$ under change of variables \eqref{b to a}:
$$
P_E \bar{a}_i = R_\w \bar{b}_i + r\w.
$$
Since $E = \{x \in \R^d :\; \< x,\w \> = r \}$, we have $P_E 0 = r \w$.
Therefore
$$
\|\bar{b}_i\| = \|R_\w \bar{b}_i\|
= \|P_E \bar{a}_i - r\w\|
= \|P_E \bar{a}_i - P_E 0\|
\le \|\bar{a}_i - 0\|
= \|\bar{a}_i\|.
$$
This proves part 2 of the lemma.
\end{proof}

\end{document}